\documentclass[a4paper,accepted=2020-04-26]{quantumarticle}
\pdfoutput=1
\usepackage[colorlinks=true
,urlcolor=blue
,anchorcolor=blue
,citecolor=blue
,filecolor=blue
,linkcolor=blue
,menucolor=blue
,linktocpage=true
,pdfa=true
]{hyperref}
\usepackage[english]{babel}
\usepackage{amsmath,amsthm,amssymb}
\usepackage{amsfonts}
\usepackage{color}
\usepackage{graphicx}
\usepackage{pgfplots}
\usepackage{tikz}
\usepackage{relsize}
\usepackage{bbm}
\usepackage{ upgreek }
\usepackage{mathrsfs}
\usepackage[numbers,compress]{natbib}
\usepackage{relsize}
\usepackage{appendix}
\usepackage{appendix}

\def\bk{{\bf k}}
\def\br{{\bf r}}

\def\bi{{\bf i}}
\def\bj{{\bf j}}
\def\bd{{\bf d}}

\newcommand{\om}{\omega}
\newcommand{\Om}{\Omega}

\newcommand{\be}{\beta}
\newcommand{\la}{\lambda}
\newcommand{\La}{\Lambda}
\newcommand{\al}{\alpha}

\newcommand{\si}{{\sigma}}

\newcommand{\del}{{\partial}}

\newcommand{\mc}[1]{\mathcal{#1}}
\newcommand{\id}{\mathbbm{1}}
\newcommand{\tr}{\mathrm{tr}}
\newcommand{\sfrac}[2]{{\textstyle\frac{#1}{#2}}}
\newcommand{\half}{\frac{1}{2}}
\newcommand{\shalf}{\sfrac{1}{2}}

\newcommand{\nn}{\nonumber}
\newcommand{\ep}{\varepsilon}

\newcommand{\dd}{\mathrm{d}}
\newcommand{\ket}[1]{\left| #1 \right>} 
\newcommand{\bra}[1]{\left< #1 \right|}

\renewcommand{\Re}{\mathrm{Re}}
\renewcommand{\Im}{\mathrm{Im}}
\newcommand{\mrm}{\mathrm}
\newcommand{\nodagger}{{\phantom{\dagger}}}
\newcommand{\GHZ}{\ket{\mathrm{GHZ}}}
\newcommand{\GHZp}{\ket{\mrm{GHZ}'}}

\newcommand{\defeq}{\mathrel{\raisebox{.034em}{:}}=}

\begin{document}
	
\title{Conditions for superdecoherence}

\author{Joris Kattem\"olle}
\orcid{0000-0003-0999-0162}

\author{Jasper van Wezel}
\orcid{0000-0002-9378-008X}

\affiliation{Institute for Theoretical Physics, University of Amsterdam,
	Science Park 904, Amsterdam, The Netherlands}
\affiliation{QuSoft, CWI,
	Science Park 123, Amsterdam, The Netherlands}

 \begin{abstract}
	Decoherence is the main obstacle to quantum computation. The decoherence rate per qubit is typically assumed to be constant. It is known, however, that quantum registers coupling to a single reservoir can show a decoherence rate per qubit that increases linearly with the number of qubits. This effect has been referred to as superdecoherence, and has been suggested to pose a threat to the scalability of quantum computation.  Here, we show that superdecoherence is absent  when the spectrum of the single reservoir is continuous, rather than discrete. The reason of this absence, is that, as the number of qubits is increased,  a quantum register inevitably becomes susceptible to an ever narrower bandwidth of frequencies in the reservoir. Furthermore, we show that for superdecoherence to occur in a reservoir with a discrete spectrum, one of the frequencies in the reservoir has to coincide exactly with the frequency the quantum register is most susceptible to. We thus fully resolve the conditions that determine the presence or absence of superdecoherence. We conclude that superdecoherence is easily avoidable in practical realizations of quantum computers.
\end{abstract}

\maketitle

\section{Introduction}\label{sec:introduction}
In principle, quantum computers can solve problems that are intractable on any classical computer. The largest obstacle to bringing this in practice is decoherence \cite{schlosshauer2007decoherence}, and it is essential to understand the sources and effects of decoherence under practical circumstances encountered in actual quantum computers. As we inch towards full-scale quantum computing, where we are already facing systems with on the order of a hundred qubits \cite{preskill2018quantum, arute2019quantum,zhang2017observation}, the system size dependence of decoherence becomes of increasing importance.
	
	Decoherence is commonly studied in a simplified spin-boson model, where only the dephasing effects of the bosonic bath are taken into account \cite{unruh1995maintaining, palma1996quantum,reina2002decoherence, breuer2002theory,leggett1987dynamcis, zanardi1997noiseless,duan1998reducing, benedetti2014effective, anton2012pure,addis2014coherence}.  Henceforth we will refer to this model as simply `the dephasing model'. This model is exactly solvable, and at the same time broadly relevant because dephasing times are typically much shorter than relaxation times  \cite{reina2002decoherence, palm2017nonperturbative, benedetti2014effective}. It should be noted, however, that there are situations where it does not accurately describe the decoherence process because of non-perturbative effects \cite{palm2017nonperturbative,benedetti2014effective}. If, in the dephasing model, each qubit is assumed to couple to its own, independent reservoir, the decoherence rate per qubit is constant. If, on the other hand, the qubits couple to single reservoir, the decoherence rate per qubit scales linearly with the number of qubits for certain states \cite{palma1996quantum, stockburger2007superdecoherence, berman2005collective, reina2002decoherence, breuer2002theory, galve2017microscopic, cirone2009collective}. This effect has been referred to as superdecoherence, in analogy with superradiance. 
	
	Superdecoherence has been predicted in Refs. \mbox{\cite{reina2002decoherence, palm2017nonperturbative}},  and has been observed experimentally in an ion-trap quantum computer \cite{monz201114}. Although some states suffer superdecoherence, the probability of running into such a state during the course of an actual algorithm may be extremely small \cite{berman2005collective}. Additionally, if the decoherence is dominated by relaxation, rather than dephasing, is has been shown that superdecoherence does not occur for the Greenberger-Horne-Zeilinger (GHZ) and the Hadamard state \cite{dalton2003scaling}. Also the particular model of solid-state qubits coupling to a single phonon reservoir has been shown not to give rise to superdecoherence  \cite{ischi2005decoherence}. The latter approach focuses on a specific setting of the dephasing model: the geometry of the quantum register is assumed to be a linear array, and the phonon reservoir is assumed to be three-dimensional and thermal, with a continuous spectrum and a linear dispersion relation. Therefore, it is unable to reveal the general underlying physical reasons for the absence of superdecoherence. The reason why superdecoherence emerges in other settings of the single-reservoir dephasing model remained unknown.    
	
	Here, we fully resolve the physical conditions that determine the presence or absence of superdecoherence in the dephasing model, with all qubits coupling to a single bath. We do not make any assumptions about the geometry of the quantum register, the dimension $d$, the reservoir dispersion relation, or the directional dependence of the spin-boson interaction. For the reservoir state, we assume a very general initial condition that applies to practically relevant situations. In this general setting, we find that the (im)possibility of superdecoherence due to a single reservoir is completely determined by the boundedness of the \emph{spectral density} and the \emph{occupation density} of the reservoir.
	
	The spectral density is the density of modes at a given frequency. If the reservoir admits only a discrete set of frequencies, such as the electromagnetic field in an ideal cavity, the spectral density is given by a sum of delta functions, and is hence unbounded. If, on the other hand, the reservoir admits a continuum of frequencies, such the electromagnetic field in an imperfect cavity or free space,  the reservoir spectral density is a bounded function of frequency.
	
	The occupation density, on the other hand, tells us to what extent a given mode in the reservoir is exited. It is typically a bounded function of the mode frequency. However, if only a single frequency is excited, the occupation density is described by a delta function which is centered at that frequency. This is the case when the bosonic field is the electromagnetic field, and a mode is excited by a laser with vanishing spectral bandwidth. In contrast, if this laser has a nonzero spectral bandwidth,  also the occupation density remains bounded. 
	
	Here, we prove that superdecoherence is absent in single-reservoir dephasing if both the reservoir spectral density and the reservoir occupation density are bounded. We call these reservoirs continuous because, in this case, both spectra have continuous support. An important physical quantity in the proof is the \emph{dephasing susceptibility}, which we define as the only part of the decoherence rate that depends on the system. It is closely related to, but different from, the so-called array factor, which arises in classical antenna arrays \cite{balanis2016antenna}, quantum antenna arrays \cite{inigo2018quantum}, and interdigital transducers that couple to surface acoustic waves \cite{morgan2010surface}. The dephasing susceptibility captures the extent to which a reservoir frequency contributes to the dephasing process if this frequency is present in the reservoir. 
	
	The reason for the presence of superdecoherence in some situations, is that there may be frequencies for which the dephasing susceptibility scales quadratically with the number of qubits. If one of these frequencies coincides with a frequency for which either the reservoir spectral density or the reservoir occupation density diverges, superdecoherence is exhibited. This is because, in this specific case, the decoherence rate scales with the system size in the same way as the peak of the dephasing susceptibility. This explains why superdecoherence is exhibited when either the spectral density or the occupation density is unbounded.
	
	The reason for the absence of superdecoherence in continuous reservoirs (i.e. bounded spectral density and occupation density) is that peaks in the dephasing susceptibility inevitably become narrower as the system size is increased.  Specifically, we show that, if the dephasing susceptibility has a peak whose height scales as the square of the number of qubits, the width of this peak must scale inversely with the number of qubits. That is to say, the quantum register may be increasingly susceptible to a given reservoir frequency as the system size grows, but the bandwidth of this susceptibility must at the same time decrease. This effect mitigates the total decoherence rate, and  the net effect is that superdecoherence is suppressed.
	
	\usetikzlibrary{arrows}
	\newcommand{\fac}{1.5}
	\begin{figure}
		\begin{tikzpicture}[xscale=\columnwidth/285,yscale=\columnwidth/285]
		\draw[smooth,variable=\x, blue!50!white, domain=0:10, thick] plot ({\x},{\fac*sin(deg(2*3.1415*\x/10))/2}) ;
		\foreach \x in {0.5,1.5,2.5,3.5,4.5,5.5,6.5,7.5,8.5,9.5} {
			\filldraw[blue!50!white] (\x,{\fac* sin(deg(2*3.1415*\x/10))/2}) circle (.075);
			\draw[-angle 60, thick] (\x,\fac*-1/3 / 2) -- (\x,\fac*1/3 / 2) ;
		}
		\draw[|<->|] (1.5,-.75) -- (2.5,-.75) node[below, pos=.5] {$a$};
		\end{tikzpicture}
		
		\vspace{.8em}
		
		\hrule

		\begin{center}
			\begin{tikzpicture}
			\begin{axis}[grid=major,trig format plots=rad, width=\columnwidth, height=1/2*\columnwidth ,xtick={-3.1415,-.6283,0,.6283,3.1415}, xticklabels={$-\frac{\pi}{a}$, $\color{blue!70!white} -\frac{2\pi}{aL}$, $0$,$\color{blue!70!white} \frac{2\pi}{aL}$, $\frac{\pi}{a}$}, ytick={0,100}, axis lines = left, domain=-3.1415:3.1415 , ymax=110]
			\addplot+[thick, mark=none, samples=100, smooth,black!90!white]{(sin(10*x/2))^2/(sin(x/2))^2};
			\end{axis}
			\draw[|<->|] (2.63,2.8) --  
 (3.95,2.8) node[above, pos=.5] {$\sim\!\!\frac{1}{aL}$};
			\draw (0,3) node {$|f(k)|^2$};
			\draw (7,0) node {$k$};
			\end{tikzpicture}
		\end{center}
	\caption{\label{fig:dipole_array}
		(\textbf{Top}) A classical analogue, where a linear array of $L=10$ classical dipoles, with lattice spacing $a$, is placed in the electromagnetic field. As a whole, the array couples strongly to the  mode with wave number $k=0$ (not shown). The array does not couple at all to modes with wave number $k=\pm 2\pi/(aL)$ (shown in blue). This is because, for these modes, all potential energies arising form the dipole-field interaction cancel exactly. (\textbf{Bottom}) The modulus squared of the coupling strength in the classical analogue, as a function of the wave number $k$, depicting two modes that do not couple to the array in blue. The array mainly couples to modes in a bandwidth less than $\Delta k = 4\pi/(a L)$, which goes as $\sim1/(aL)$. 
	}
\end{figure}
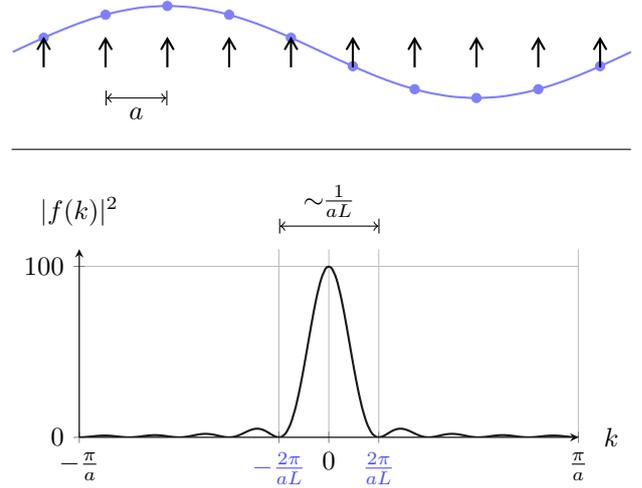
	
\subsection{A classical analogue }
The cause of the inverse scaling of the bandwidth of the susceptibility, which is responsible for the absence of superdecoherence in continuous reservoirs, can be sketched with a classical analogue. We leave the treatment of the quantum dephasing susceptibility for Sec.~\ref{sec:dephasing_susceptibility}. Consider $L$ identical, classical, noninteracting electric  dipoles in a linear array with spacing~$a$, as depicted in Fig.~\ref{fig:dipole_array} (top). (This geometry is chosen for explanatory reasons. Our results concerning the quantum dephasing susceptibility hold for general register geometries.) In the initial state of the array, all dipoles point upwards. For simplicity, consider only the electromagnetic modes whose momentum is colinear with the array and are polarized in the direction of the dipole moments. The dipoles couple to the electromagnetic field, giving an initial potential energy $V=C_1 \sum_{\ell=1}^{L} E_\ell$, where $C_1$ is some constant, $E_\ell$ is the electric field at the $\ell$th dipole, and $k$ the wave number. In terms of the Fourier transform $E(k)\defeq \sum_{\ell=1}^{L} e^{ikr_\ell}E_\ell$, where $r_\ell=a(\ell-1)$ is the position of the $\ell$th qubit, the initial  potential energy equals $V=\frac{a C_1}{2\pi} \int_{-\pi/a}^{\pi/a} \dd k\, f(k) E(k)$, with $f(k)= \sum_{\ell=1}^{L} e^{-ikr_\ell}$ the coupling strength between the array and the mode with wave number $k$. See Fig.~\ref{fig:dipole_array} (bottom) for a plot of  $|f(k)|^2$. From the previous expression for $f(k)$, and from the plot, we can see the array couples most strongly to the electromagnetic field mode with wave number $k=0$. We can also see that the array does not couple at all to modes with wave number $\pm 2\pi/(aL)$. In real space, this is because, for this wave number, all potential energies cancel exactly [also see Fig.~\ref{fig:dipole_array} (top)]. Thus, the bandwidth of modes to which the array couples strongly is at most $\Delta k = 4\pi/(aL)$, which scales inversely with the length of the array.

	\section{Spin-boson dephasing}\label{sec:pure_dephasing}
	In this section, we introduce the model of spin-boson dephasing, following references \cite{ palma1996quantum, reina2002decoherence,breuer2002theory}. First, we consider the case of a single qubit coupling to a bosonic reservoir, and extend this to multiple qubits, each of which couples to its own, independent, bosonic reservoir. In both of these cases, superdecoherence cannot occur under any circumstance. Subsequently, this situation is contrasted with the scenario where all qubits couple to a single bosonic reservoir, in which case superdecoherence may in fact occur. We make some generalizations concerning the initial reservoir state, the details of which can be found in Appendix  \ref{sec:general_reservoir_state}. We use units where $c=\hbar=k_B=1$. 
	
	\subsection{Single qubit}
	Consider a single qubit (`the system'), with an internal Hamiltonian $H_S=\Delta J^z$, that is placed in a bosonic reservoir. Here $\Delta$ is the level spacing and $J^z$ the spin-$z$ operator. We work in the computational basis, where this operator is diagonal, and has eigenstates $\ket{1/2}$ and $\ket{-1/2}$. The internal Hamiltonian of the reservoir is given by $H_B=\sum_{\bk}\omega_\bk N_\bk$, with $N_\bk = a_\bk^\dagger a_\bk^\nodagger$ the number operator of a bosonic mode with wave vector $\bk$.  Here $a_\bk^{\nodagger}$ $(a_\bk^\dagger)$ is the bosonic annihilation (creation) operator of the mode with wave vector $\bk$. The sum is over all $\bk$ that are admitted by the reservoir. The set of $\bk$s that are admitted by the reservoir depends on the physical details of the reservoir. The reservoir couples to the qubit via the interaction term $H_{SB}=\sum_{\bk} J^z (g_{\bk}^*a_\bk^{\nodagger}+g_{\bk}^\nodagger a_\bk^\dagger)$, with $g_\bk^\nodagger$ the coupling strength between the qubit and the mode with wave vector $\bk$. There are many explicit physical settings that may lead to this interaction term \cite{doll2008decoherence}, but here, we do not assume such a specific setting. Since the only system operator in the interaction term is $J^z$, $H_{SB}$ causes dephasing only. Putting all terms together, the dephasing model of a single qubit reads
	\begin{equation*}
		H_1\defeq\Delta J^z+\sum_{\bk} \omega_\bk N_\bk+\sum_{\bk} J^z (g_{\bk}^*a_\bk^{\nodagger}+g_{\bk}^\nodagger a_\bk^\dagger).
	\end{equation*}
	
	In this and the following sections, we assume that the overall system-reservoir state is a product state, $\rho(0)\otimes \rho_B(0)$. Here $\rho(0)$  (no subscript) is a general initial system state, and $\rho_B(0)$ is the initial reservoir state. The latter is assumed to be a product state of single-mode states, $\rho_B(0)=\bigotimes_\bk \rho_{B,\bk}(0)$, with $\rho_{B,\bk}(0)$ the initial state of the mode with wave vector $\bk$. The state $\rho_{B,\bk}(0)$ is assumed to be a displaced thermal state, that is, $\rho_{B,\bk}(0)=D(\alpha_\bk) e^{-\om_\bk N_\bk /T_\bk} D^\dagger(\al_\bk)/\mc Z$, where $\al_\bk$ is the displacement (which can be any complex number),  $N_\bk$ the number operator, $T_\bk$ the  ($\bk$-dependent) temperature, $\mc Z$ the normalization, and $D$ the displacement operator. (In App.~\ref{sec:general_reservoir_state} we show displacement is irrelevant in the dephasing process, so we do not give an expression for $D$ here.) Possible $\rho_{B,\bk}(0)$ admitted by this parameterization include the regular single-mode thermal states ($T_\bk\geq 0$ and $\al_\bk=0$), the coherent states ($T_\bk=0$, $|\al_\bk|\geq 0$), and the vacuum state ($T_\bk=0$, $\al_\bk=0$). We call a reservoir \emph{completely thermal} if the overall initial reservoir state $\rho_B(0)$ equals the regular thermal density matrix with temperature $T$, that is, if  $\rho_B(0)=e^{-\om_\bk N_\bk/T}/\mc{Z'}$. In our parameterization of initial reservoir states, this is the specific case where $\alpha_\bk=0$ and $T_\bk=T$ for all $\bk$. Our form of the initial reservoir state is a generalization of that used in references \cite{unruh1995maintaining, palma1996quantum, reina2002decoherence,breuer2002theory, duan1998reducing,addis2014coherence,berman2005collective}, where the assumption is that the initial reservoir state is completely thermal.
	
	It can be shown that the absolute value of the $i,j$th entry (with $i,j\in\{-1/2,1/2\}$) of the system density matrix, after time $t$, is given by
	\begin{equation}\label{eq:single_qubit}
		\lvert \rho_{ij}(t)\rvert = e^{-\Gamma_{i-j}(t)} \lvert \rho_{ij}(0)\rvert,
	\end{equation}
	where $\Gamma_{i-j}(t)$ is  the \emph{decoherence function} (refs. \mbox{\cite{ palma1996quantum, reina2002decoherence,breuer2002theory}},\footnote{These references give a derivation for the density operator in the interaction picture, $|\rho^{Int}_{i j}(t)|=e^{-\Gamma_{i-j}}|\rho^{Int}_{ij}(0)|$. In the dephasing model, $|\rho^{Sch}_{ij}(t)|=|\rho^{Int}_{ij}(t)|$.  Therefore, we drop the superscript indicating the picture in Eq.~(\ref{eq:single_qubit}), keeping in mind that the equation holds in both pictures. The same applies to the system density operators in Sec.~(\ref{sec:independent_pure_dephasing}) and (\ref{sec:collective_pure_dephasing}).} App.~\ref{sec:general_reservoir_state}). In the current model, dephasing is the only decoherence mechanism. Therefore the  \emph{decoherence rate} can be defined as $1/T_2$,  were $T_2$ is the dephasing time,  here defined as the smallest time $t$ for which $\Gamma_{i-j}(t)=1$. 
	
	In general, the decoherence function only depends on the difference $\dd=i-j$.\footnote{We use the italic $d$ for dimension, and the straight $\mrm d$ for the differences $\mrm d=i-j$ and (for multiple qubits) $\bd=\bi-\bj$.} It is given by
	\begin{equation}\label{eq:Gamma_single_qubit} \Gamma_\mrm{d}(t)=\gamma_\mrm{d}\sum_{\mathbf{k}} |g_\mathbf{k}|^2\tau(t,\om_\bk)(1+2 \bar N_\bk).
	\end{equation}
	Here $\gamma_\dd=|\dd|$ and 
	\begin{equation}\label{eq:tau}
		\tau(t,\om_\bk)=\frac{1-\cos(\omega_\bk t)}{\om_{\bk}^2}.
	\end{equation}
	Under the current assumptions on the initial reservoir state, the occupation number $\bar N_\bk$ of the mode $\bk$ is given by the Bose-Einstein distribution with \mbox{($\bk$-dependent)} temperature $T_\bk$.\footnote{The initial state of the reservoir may still be a general thermal \emph{displaced} state. Displacement of a mode does affect the expectation value of its number operator, but only the thermal part contributes to $\Gamma_\bd(t)$. See Appendix~\ref{sec:general_reservoir_state} for details.} That is,
	\begin{align}\label{eq:Bose-Einstein}
		\bar N_\bk=\frac{1}{e^{\om_\bk/T_\bk}-1}.
	\end{align}
	This need not be an isotropic function on $k$-space. For the specific case of the completely thermal reservoir (i.e. $T_\bk=T$ and $\al_\bk=0$ for all $\bk$), the occupation number is in fact isotropic, and depends on the mode energy only,
	\begin{equation}\label{eq:thermal_spectrum}
		\bar N_\om^{th} \defeq \frac{1}{e^{\om/T}-1}.
	\end{equation}
	We do not assume any particular dispersion relation, nor the reservoir to be completely thermal, unless stated otherwise. 
	
	\subsection{Independent reservoirs}\label{sec:independent_pure_dephasing}
\begin{figure}
	\centering
	\includegraphics[width=1\columnwidth]{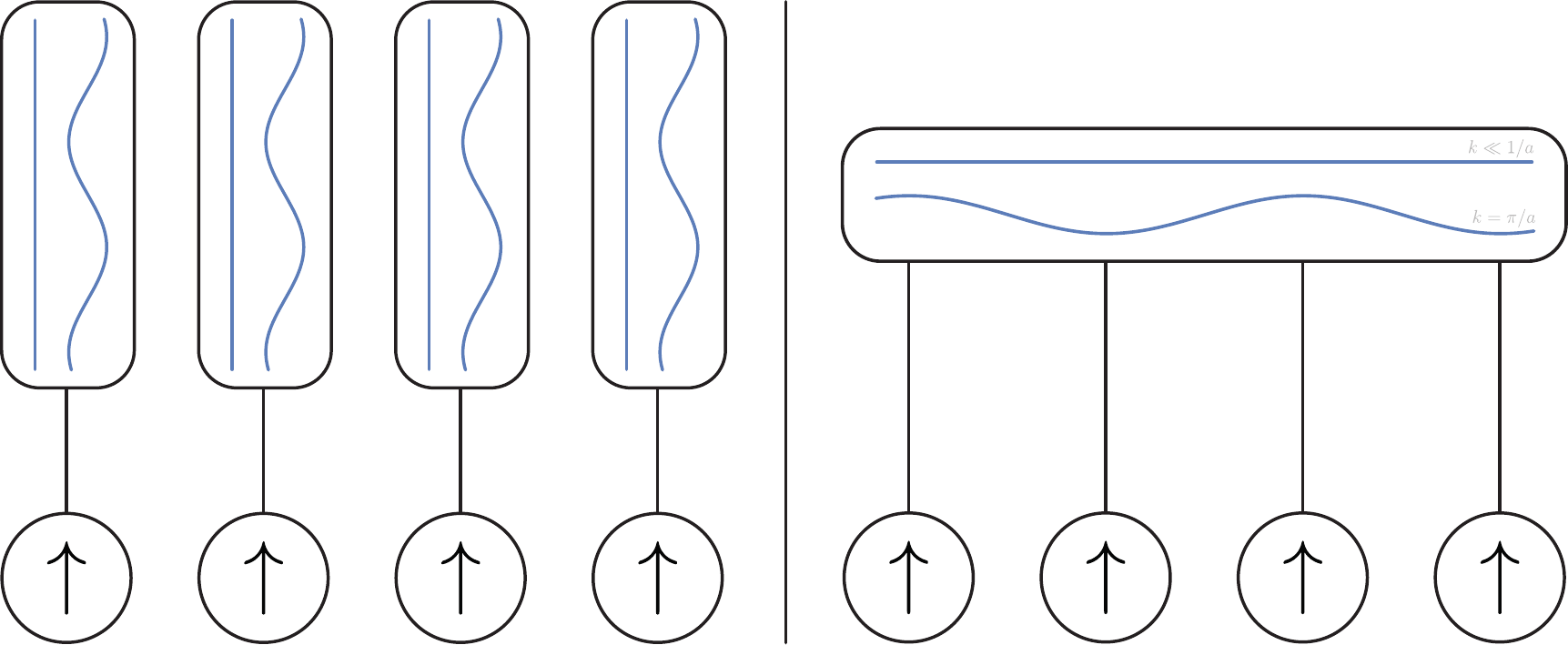}	
	\caption{\label{fig:independent_vs_collective} \textbf{(Left)} Qubits coupling to independent reservoirs. \textbf{(Right)} Qubits coupling to a single reservoir. }
\end{figure}

	Now consider $L$ copies of the system-reservoir combination described in the previous subsection. This setting is known as independent dephasing. The overall Hamiltonian reads $H_L^\mrm{ind}=(H_1)^{\otimes L}$. This is depicted schematically in Fig.~\ref{fig:independent_vs_collective} (left).
	
	We denote states in the computational basis of the $L$-qubit quantum register by $\ket \bi \equiv \ket{i_1,\ldots,i_L}$. It can be shown that, under the evolution by $H^{\mrm{ind}}_L$, the absolute value of the $(\bi,\bj)$th entry of the system density matrix equals $
		\lvert\rho_{\bi\bj}(t)\rvert=e^{-\Gamma_{\bd}(t)}\lvert\rho_{\bi\bj}(0)\rvert,$
	with $\bd$ the difference vector $\bd=\bi-\bj$ and
	\begin{equation*}
		\Gamma_{\bd}(t)=\gamma_{\bd}\sum_{\bk} |g_\mathbf{k}|^2 \tau(t,\om_\bk) (1+2\bar N_\bk)
	\end{equation*}
	the decoherence function. Here, we have singled out the factor $\gamma_{\bd}=\sum_{\ell=1}^L|\bd_\ell|$ for later reference. This factor is the only part of the decoherence function that depends on $L$, and it is at most proportional to $L$. Thus, for independent dephasing, the decoherence function scales at worst linearly with the system size,
	\begin{equation*}
	\Gamma_\bd \propto L.
	\end{equation*}
	That is, the decoherence rate per qubit is at most constant in the system size.
	
	\subsection{A single reservoir}\label{sec:collective_pure_dephasing}
	Now consider the situation where all qubits couple to a single reservoir,
	\begin{align}\label{eq:collective_pure_dephasing}
		H=&\Delta\sum_{\ell=1}^{L} J_\ell^z+\sum_{\bk} \omega_\bk N_\bk\nn\\
		&+\sum_{\ell=1}^{L} \sum_{\bk} J_\ell^z (g_{\bk \ell}^*a_\bk^{\nodagger}+g_{\bk \ell}^\nodagger a_\bk^\dagger),
	\end{align}
	as is depicted schematically in Fig.~(\ref{fig:independent_vs_collective}).
	Again, $\bk$ runs over all wave vectors that are supported by the reservoir. Now, the coupling constant $g_{\bk \ell}$ depends on both the wave vector and the qubit location. If the reservoir consist of plane-wave modes, $g_{\bk \ell}=g_{\bk}e^{i \bk \cdot \br_\ell} $.
	For single-reservoir dephasing, it can be shown that  (Refs. \cite{ palma1996quantum, reina2002decoherence,breuer2002theory}, App. \ref{sec:general_reservoir_state}) the density matrix equals
	\begin{equation*}
		|\rho_{\bi\bj}(t)|= e^{-\Gamma_{\bi - \bj}(t)}\lvert\rho_{\bi\bj}(0)\rvert,
	\end{equation*}
	as before, but now
	\begin{equation}\label{eq:Gamma_sum}
		\Gamma_\bd(t)=\sum_{\bk}\gamma_\bd(\bk) |g_\mathbf{k}|^2 \tau(t,\om_\bk)(1+2\bar N_\bk),
		\end{equation}
	with
	\begin{equation}\label{eq:gamma}
		\gamma_\bd(\bk)=\sum_{\ell m}\bd_\ell \bd_m \cos(\bk\cdot \br_{\ell m}),
	\end{equation}
	where $\br_{\ell m} \defeq \br_\ell - \br_m$ is the vector pointing from the location of qubit $\ell$ to that of qubit $m$.	In contrast to the situation of independent dephasing, $\gamma_{\bd}(\bk)$ now depends on $\bk$ and contains a double sum over the qubit indices. The summand of $\gamma_{\bd}(\bk)$ can at most equal unity, which is attained, for example, if $\bd_\ell = 1$ for all $\ell$, and $\bk=\mathbf 0$. Thus, if indeed $\bk=\mathbf{0}$ is admitted by the reservoir,
	\begin{equation*}
	\Gamma_\bd \propto L^2
	\end{equation*}
	at worst. The possibility of quadratic, rather than linear scaling of the decoherence function with $L$ is called \emph{superdecoherence}. The decoherence rate per qubit can thus scale with the system size, which is problematic for error correction \cite{gottesman1997stabilizer, preskill1998lecture}.
	
	\subsection{The continuum limit}
	We may write Eq.~(\ref{eq:Gamma_sum}) in a more meaningful form, starting by introducing ${\mc D} = \sum_{\bk'}\delta(\bk-\bk')$, so that we may replace the sum by an integral,
	\begin{equation*}
	\sum_{\bk}\ldots \rightarrow \int_{\mathbb R^{d}}\, \dd\bk\, {\mc D}(\bk)\ldots,
	\end{equation*}
	where $d$ is the dimension of the reservoir. Here $\mc D(\bk)$ is a density of states on $k$-space, currently describing a discrete set of modes. Note that $\mc D$ is unbounded at those modes, and vanishes elsewhere. In the continuum limit, the peaks merge into a bounded and continuous density of states on $k$-space.  We then have
	 \begin{equation}\label{eq:Gamma_integral}
		\Gamma_\bd(t)=\int_{\mathbb R^d} \dd\bk\, \mc D(\bk) \lvert g_\bk \rvert^2 \, \gamma_\bd(\bk) \, \tau(t,\om_\bk)\, (1+2\bar N_\bk),
	\end{equation}
	where $\mc D$ is unbounded for discrete reservoirs, and bounded in the continuum limit. In the continuum limit, $\bar N_\bk$ becomes an occupation density rather than an occupation number. Note $\mc D(\bk)$ is different from the usual density of states, because the latter is a function of frequency only. For the electromagnetic field in free space, without boundary conditions, $\mc D$ is proportional to a constant with length dimension $d$. Equation (\ref{eq:Gamma_integral}) is the most general form of the decoherence function in the dephasing model because it can describe both discrete and continuous reservoirs. We will work with this form from now on. 
	
	One feature of Eq.~(\ref{eq:Gamma_integral}) (and the preceding, less general forms) is that we can easily separate the vacuum contributions form those that are due to reservoir excitations. That is, we may write
	\begin{equation}\label{eq:Gamma_separated}
	\Gamma_\bd(t) =:\Gamma^{(vac)}_\bd(t)+\Gamma^{(ex)}_\bd(t),
	\end{equation}
	with
	\begin{equation}\label{eq:Gamma_vac_ex} \Gamma^{(vac/ex)}_\bd(t)\defeq \int_{\mathbb R^d}\, \dd \bk\, \gamma_\bd(\bk)\xi^{(vac/ex)}(t,\bk),
	\end{equation}
	where
	\begin{align}
		\xi^{(vac)}(t,\bk)&\defeq \mc D(\bk)\lvert g_\bk\rvert^2 \tau(t,\om_\bk), \label{eq:xi_vac}\\
		\xi^{(ex)}(t,\bk)&\defeq \mc D(\bk)\lvert g_\bk\rvert^2 \tau(t,\om_\bk)\,2\bar N_\bk. \label{eq:xi_ex}
	\end{align}
	
	For the dephasing susceptibility to be well-defined, the integral in Eq.~(\ref{eq:Gamma_integral}) has to converge. This is guaranteed by a high frequency cutoff. Physically, this arises because, as a function of $\om_\bk$, either $\mc D$ goes to zero, or the coupling strength $g_\bk$ goes to zero, or a combination of both. Here, we assume that after some cutoff frequency $\om_c$, the product $\mc D(\bk)\lvert g_\bk\rvert^2$ is suppressed at least exponentially, \begin{equation}\label{eq:cutoff}
	\mc D(\bk)\lvert g_\bk\rvert^2=O (e^{-\om_\bk/\om_c}).
	\end{equation}  At this point,  this cutoff does not impose any restriction on the physical systems described because $\om_c$ can be arbitrarily large. 	 	
	
	Even in continuous reservoirs, it is possible in theory that a single mode $\bk'$ is excited, but no modes in its neighborhood (in $k$-space). Then, the occupation density is unbounded at that mode, $\bar N_\bk \propto \delta(\bk-\bk')$. We call a reservoir \emph{continuous} if, in contrast, both $\mc D(\bk) \lvert g_\bk \rvert^2$  and $\bar N_\bk$ are bounded functions of $\bk$.
	
	A common assumption \cite{palma1996quantum, reina2002decoherence,breuer2002theory}, that we will only make occasionally, is that $\mc D(\bk) \lvert g_\bk \rvert^2$ and $\bar N_\bk$ are isotropic, and that the reservoir dispersion relation is linear. For a linear dispersion relation, $\om_\bk= v |\bk|$ for some constant $v$. Working in units where $v=1$ for notational convenience, we may then transform to spherical coordinates and write
	\begin{equation}\label{eq:isotropic}
	\Gamma_\bd(t)= \int_0^{\infty} \dd \om \, J(\om) \tilde \gamma_\bd(\om) \tau(t,\om)(1+2\bar N_\om),
	\end{equation}
	with
	\begin{equation*}
	\tilde\gamma_\bd(\om) \defeq \int \dd \Om \, \gamma_\bd(\om,\theta).
	\end{equation*}
	Here $\Om$ is the $d-1$ dimensional solid angle, and $\theta$ the $d-1$ dimensional angle of $\bk$. The function $J(\om)=\om^{d-1}\mc D(\om)|g_\om|^2$ is called the \emph{spectral density} of the reservoir. A common form is \cite{unruh1995maintaining, palma1996quantum, reina2002decoherence,breuer2002theory, leggett1987dynamcis, benedetti2018quantum,bulla2003numerical,vojta2005quantum,anders2007equilibrium,kehrein1996on,addis2014coherence}
	\begin{equation}\label{eq:spectral_density}
	J(\omega)=\alpha_d \omega^d e^{-\om/\om_c},
	\end{equation}
	with $\al_d$ a constant with length dimension $d-1$, and $\om_c$ the cutoff frequency. This expression is often extended to include even non-integer $d$, which may be encountered in reservoirs with fractal properties \cite{leggett1987dynamcis}. Depending on the dimension, these reservoirs are called subohmic ($d<1$), Ohmic ($d=1$), or superohmic ($d>1$). In this manuscript, we do not assume isotropy, unless stated otherwise, and we will manly work with the general form of the decoherence function [Eq. (\ref{eq:Gamma_integral})].
	
	In the following sections, we study the qualitative system-size scaling of the decoherence function.  For completeness, however, in  Appendix~\ref{sec:explicit_expressions_for_the_vacuum_contribution} we show  explicit solution for $\Gamma^{(vac)}_L$, and derive simplified approximate solutions in the regimes $t\ll 1$ and $t\to\infty$.
	
	\section{Dephasing susceptibility}\label{sec:dephasing_susceptibility}
	In this section, we identify $\gamma_\bd(\bk)$ as an important physical quantity and derive some of its properties, especially regarding its system size dependence. Namely, $\gamma_\bd(\bk)$ is determined solely by the system, and it is the only part of the decoherence function that depends on the system. So it fully captures the influence of the system on the decoherence function. The function $\gamma_\bd(\bk)$ weighs the severity of the influence of the mode $\bk$ if this mode was to be `offered' by the reservoir, and depends on the system geometry and the index $(\bi,\bj)$. We call it the \emph{dephasing susceptibility} of the reservoir. In Appendix~\ref{sec:dynamical_fidelity_susceptibility_of_pure_dephasing}, we show how this susceptibility relates to the dynamical fidelity susceptibility of decoherence-free subspaces, which we introduced in previous work \cite{kattemolle2019dynamical}.
	
	To illustrate the qualitative behavior of the dephasing susceptibility, we first consider the \emph{array model}. It consists of a linear array of $L$ noninteracting qubits with spacing $a$ that couple to a single reservoir with dimension $d=1$. Two system states we consider are 
	\begin{equation}\label{eq:GHZ}
	\begin{aligned}
		\GHZ&=\sfrac{1}{\sqrt 2}\ket{\shalf,\shalf}^{\otimes L/2}+\sfrac{1}{\sqrt 2}\ket{-\shalf,-\shalf}^{\otimes L/2},\\
		 \GHZp&=\sfrac{1}{\sqrt 2}\ket{\shalf,-\shalf}^{\otimes L/2}+\sfrac{1}{\sqrt 2}\ket{-\shalf,\shalf}^{\otimes L/2}.
		\end{aligned}
	\end{equation}
	Both states are of the form $(\ket\bi+\ket\bj)/\sqrt 2$, and thus have only a single nonzero matrix element in the upper right triangle of their density matrix. That is, in the computational basis,
	\begin{equation}\label{eq:GHZmatrix}
	\rho_{\mrm{GHZ}}=\frac{1}{2}
	\left(
	\begin{array}{ccccc}
	1			&	0	&	\ldots	&	0	&	1 \\
	0			&	0	&	\ldots	&	0	&	0 \\
	\vdots	&	\vdots	 &	\ddots & 	  \vdots    & \vdots \\
	0			&	0	&	\ldots	&	0	&	0 \\
	1			&	0	&	\ldots	&	0	&	1 \\
	\end{array}
	\right),
	\end{equation}
	and similarly for the density matrix associated with~$\GHZp$.
	
\begin{figure}
		\centering
		\begin{tikzpicture}
		\begin{axis}[trig format plots=rad, cycle list name=exotic, width=\columnwidth, xtick={0,3.1415,6.289}, xticklabels={0, $\frac{\pi}{a}$, \hspace{-1em} $\frac{2\pi}{a}$}, ytick={0,20,40,60},axis lines = left, title=Dephasing susceptibility,xlabel={\raisebox{-1.5em}{$k$}}, ylabel=$\gamma_{\mrm{GHZ}^{'}}(k)$, domain=0.001:6.289, ymax=70]
		\addplot+[thick, mark=none, samples=100, smooth]{(sin(8*x/2))^2/(cos(x/2))^2};
		\addlegendentry{$L=8$}
		\addplot+[thick, mark=none, samples=50, smooth]{sin(6*x/2)^2/cos(x/2)^2};
		\addlegendentry{$L=6$}
		\addplot+[thick, mark=none, samples=50, smooth]{sin(4*x/2)^2/cos(x/2)^2};
		\addlegendentry{$L=4$}
		\addplot+[thick, mark=none, samples=50, smooth]{(sin(2*x/2))^2/(cos(x/2))^2};
		\addlegendentry{$L=2$}
		\end{axis}
		\end{tikzpicture}
		\caption{\label{fig:susceptibility} The dephasing susceptibility of the off-diagonal matrix element of the state $\GHZp$, for system sizes $L=2,4,6,8$. The peaks have hight $L^2$ and width $\sim 1/L$. For the off-diagonal matrix element of the state $\GHZ$, the entire graph is translated in such a way that the peaks lie above ${k=0}$.}
	\end{figure}
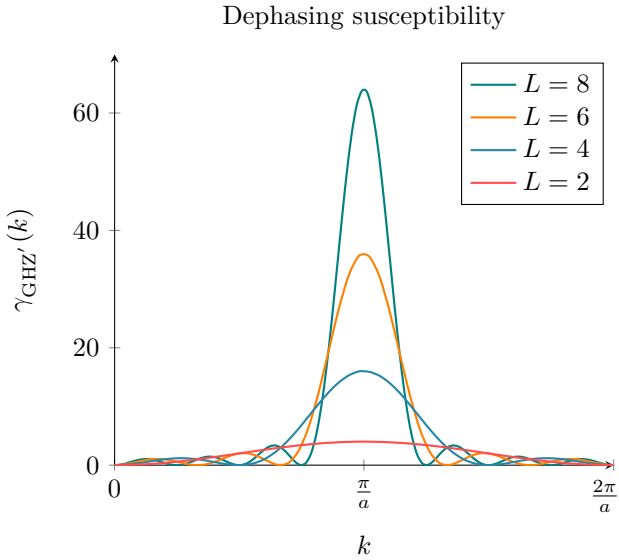

	 The difference vectors $\bd=\bi-\bj$ belonging to these off-diagonal matrix elements are
	\begin{align}\label{eq:d_FM_AFM}
		\bd_\mrm{GHZ}&=(1,1,1,1,\ldots),\\ \bd_{{\mrm{GHZ}^{'}}}&=(1,-1,1,-1,\ldots).
	\end{align}
	Thus, with Eq.~(\ref{eq:gamma}), we find\footnote{These closed form formulas are ill-defined when the denominator vanishes. The original form [Eq.~(\ref{eq:gamma})] does not have this anomaly. It is to be understood that at these points, the closed form formulas are determined by their limit values. Then the resulting functions are smooth.}
	\begin{align}\label{eq:closed_forms}
	\gamma_{\mrm{GHZ}}(k)&=\frac{\sin^2(akL/2)}{\sin^2(ak/2)},\\
		\gamma_{\mrm{GHZ}^{'}}(k)&=\frac{\sin^2(akL/2)}{\cos^2(ak/2)},
	\end{align}
	for $L$ even. Here, we write $\gamma_\mrm{GHZ}$ instead of $\gamma_{\bd_\mrm{GHZ}}$ for conciseness, and similarly for $\gamma_{\mrm{GHZ}^{'}}$. 
	Plots of $\gamma_{\mrm{GHZ}^{'}}(k)$ for various $L$ can be found in Fig.~\ref{fig:susceptibility}.
	
	Note there are values of $k$ for which ${\gamma_{\mrm{GHZ}^{'}}(k)=0}$. This occurs when $\sin^2(akL/2)=0$ but ${\cos^{2}(ak/2)\neq 0}$. That is, when $ak=\pi+n\,2\pi/L$ for integer values of $n$, excluding $n$ that are multiples of  $L/2$. (I.e. $n\in \mathbb Z \setminus \{m\in\mathbb Z\, | \, m=\ell L/2 \wedge\ell \in\mathbb Z\}$.) If the reservoir only supports these modes, the off-diagonal matrix element of $\GHZp$ does not diminish as a function of time at all. In this situation the two basis states that compose $\GHZp$ [Eq.~(\ref{eq:GHZ})] are in the same decoherence-free subspace \cite{palma1996quantum, duan1997preserving, duan1998reducing, zanardi1997error, zanardi1997noiseless, lidar1998decoherence, lidar2003decoherence, kempe2001theory}.
	
	The dephasing susceptibility $\gamma_{\mrm{GHZ}^{'}}$ is dominated by the peak at $ak=\pi$, whose height is $L^2$. Depending on the reservoir, this may result in superdecoherence. From Eq.~(\ref{eq:Gamma_sum}), we see that if the reservoir is discrete and supports the mode $ak=\pi$, the decoherence function scales as $L^2$, even in the vacuum. We stress that, as shown by this simple example, superdecoherence is possible even when the coupling constants $g_{\bk \ell}$ depend on the the qubit location. Hence permutation symmetry of the Hamiltonian is not a prerequisite for superdecoherence.   
	
	Consider the two points around the peak where ${\gamma_{\mrm{GHZ}^{'}}=0}$. The previous equations about the minima show that the distance between these points equals $\Delta k = 4\pi/(aL)$. Thus, the bandwidth of modes the off-diagonal matrix element of $\GHZp$ is most susceptible to scales inversely with the system size. 
	
	This decreasing bandwidth is shown by the dephasing susceptibility in general. This is because, mathematically, $\gamma_\bd(\bk)$ is the spectral density of the difference vector $\bd$. That is, 
	\begin{align}
		\gamma_\bd(\bk)&=\sum_{\ell m}\bd_\ell \bd_m \{\cos(\bk\cdot \br_{\ell m})+i\sin(\bk\cdot \br_{\ell m})\}\nn\\
		&=\sum_{\ell m}\bd_\ell\bd_m e^{i\bk\cdot \br_{\ell m}}\nn\\
		&=|\tilde{\mrm{d}}(\bk)|^2, \label{eq:gamma_spectral_density}
	\end{align}
	with $\tilde{\mrm{d}}(\bk) \defeq \sum_{\ell=1}^{L} e^{-i\bk\cdot\br_\ell} \bd_\ell$ the Fourier transform of $\bd$.
	Here the sine vanishes because it is antisymmetric under exchange of $\ell$ and $m$. If the qubits are placed on a lattice, the dephasing susceptibility is periodic in $\bk$.
	
	The summand in the original definition of $\gamma_\bd(\bk)$ [Eq. (\ref{eq:gamma})] is at most unity. This is achieved, for example, when $\bd_n=1$ for all $n\in\{1,\ldots,L\}$ and $\bk=0$. Thus,
	\begin{equation}\label{eq:gamma_leq_L2}
	0\leq \gamma_{\bd}(\bk)\leq L^2.
	\end{equation}
	Nevertheless, the integral of the dephasing susceptibility over one reciprocal unit cell $\mc C$ is bounded by $2\pi L/V$, where $V$ is the volume one real-space unit cell. This follows directly from the fact that the dephasing susceptibility is the spectral density function of $\bd$, and Parseval's theorem,
	\begin{align*}
		\int_{\mc C} \dd\bk\, \gamma_L(\bk)&=\int_{\mc C}\dd\bk\,\lvert \tilde{\mrm{ d}}(\bk)\rvert^2\\
		&=\frac{2\pi}{V} \sum_{m=1}^L \lvert \bd_m \rvert^2,
	\end{align*}
	with $\lvert \bd_m \rvert={\lvert\bi_m-\bj_m\rvert\leq 1}$. Therefore
	\begin{equation}\label{eq:parseval}
	\int_{\mc C} \dd\bk\, \gamma_L(\bk) \leq \frac{2\pi L}{V}.
	\end{equation}
	This shows that if the dephasing susceptibility has a peak of height $L^2$, the width of that peak must scale as $1/L$. As we show in the following section, this relation causes a mitigation of the dephasing process, causing the absence of superdecoherence in continuous reservoirs. 
	
	A related question about the dephasing susceptibility is how large $\gamma_{\bi-\bj}(\bk)$ is typically if we fix $\bk$ and $L$ and vary $(\bi,\bj)$. In Appendix~\ref{sec:typical_dephasing_susceptibility}, we show that the distribution of $\gamma_{\bi-\bj}(\bk)$ over $(\bi,\bj)$ is approximated by a Gaussian, with a standard deviation that is at most $L/(2\pi)$. This means $\gamma_{\bi-\bj}(\bk)$ is typically on the order of $L$ and that there are few $\bi-\bj$ such that  $\gamma_{\bi-\bj}(\bk)\approx L^2$.
	\section{Asymptotic system size scaling}\label{sec:system_size_scaling}
	In this section, we derive our main results, which are upper bounds on the system size scaling of the decoherence function. An important quantity herein is the decoherence function, because this is the only factor in the integrand of the  decoherence function that depends on the system size. In turn, the dephasing susceptibility depends on $L$ because $L$ is the length of the vector $\bd$. The exact scaling of $\Gamma_\bd$ with $L$ depends on how entries are added to $\bd$ as the $L$ is increased. In principle, this can be done according to any prescription. 
	
	For example, we could consider the dephasing associated with $\bd$, where $\bd$ increases in length by adding a random number for every qubit we add. A more physically relevant situation, is for example to consider the coherence of the state $\GHZ$ or $\GHZp$, as a function of the system size. The results in this section hold for any description, unless stated otherwise, but some descriptions may arise more naturally than others.
	
	To tidy up notation, and to highlight $L$ dependence, we will now write $\gamma_L(\bk)$ instead of $\gamma_{\bd}(\bk)$ and likewise $\Gamma_L(t)$ instead of $\Gamma_{\bd}(t)$. At the same time, we use $\gamma_\mrm{GHZ}(\om)$ and $\gamma_{\mrm{GHZ}^{'}}(\om)$ for the dephasing susceptibilities of the off-diagonal matrix elements of $\GHZ$ and $\GHZp$, respectively. Likewise, we write $\Gamma_\mrm{GHZ}$ and $\Gamma_{\mrm{GHZ}^{'}}$.
	
	The starting point of our derivation is the most general form of the decoherence function, in which the vacuum contributions are separated from the excitation contributions [see Eq.~(\ref{eq:Gamma_separated})]. Both contributions are of the form of Eq.~(\ref{eq:Gamma_vac_ex}). Assume $t=t_0$ is fixed. The mathematical property that the integral of $\gamma_L$ over one reciprocal unit cell scales linearly with the number of qubits [Eq.~(\ref{eq:parseval})] ensures that also $\Gamma_L^{(vac/ex)}$ scales linearly with the number of qubits, provided that $\xi^{(vac/ex)}$ is bounded.
	
	This is shown as follows. Assume $\xi^{(vac/ex)}(t_0,\bk)$ is bounded. The integral $\Gamma^{(vac/ex)}_L$ equals a sum of integrals, where each domain of integration is one reciprocal unit cell $\mc C$,
	\begin{equation*}
	\Gamma_L^{(vac/ex)}(t_0) = \sum_{\mc C}\int_{\mc C}\dd\bk\, \gamma_L(\bk) \xi^{(vac/ex)}(t_0,\bk).
	\end{equation*}
	Each term is upper bounded by the integral of $\gamma_L(\bk)$ over a single reciprocal unit cell after the integral is rescaled by the maximum of $\xi^{(vac/ex)}(t_0,\bk)$ on that unit cell,
	\begin{equation*}
		\Gamma_L^{(vac/ex)}(t_0)  \leq \sum_{\mc C} \max_{\bk\in \mc C}\left[\xi^{(vac/ex)}(t_0,\bk)\right] \int_{\mc C}\dd \bk\, \gamma_L(\bk).
	\end{equation*}
	By Eq.~(\ref{eq:parseval}),
	\begin{align*}
		\Gamma_L^{(vac/ex)}(t_0) \leq \frac{2\pi L}{V}  \sum_{\mc C} \max_{\bk\in \mc C}\left[\xi^{(vac/ex)}(t_0,\bk)\right].
	\end{align*}
	The high-frequency cutoff [Eq.~(\ref{eq:cutoff})] ensures the sum converges, no matter the value of the cutoff $\om_c$. Thus, we obtain the main mathematical result of this paper: if $\xi^{(vac/ex)}(t_0,\bk)$ is bounded, then
	\begin{equation}\label{eq:no_superdecoherence}
	\Gamma^{(vac/ex)}_L=O(L).
	\end{equation}
	
	The relevant physical question then, is when $\xi^{(vac/ex)}(t_0,\bk)$ is bounded. First, consider the vacuum contribution $\xi^{(vac)}(t_0,\bk)=\mc D(\bk)\lvert g_\bk \rvert^2\tau(t_0,\om_\bk)$ [Eq.~(\ref{eq:xi_vac})]. The temporal factor $\tau(t_0,\om_\bk)$ is a bounded function of $\om_\bk$ for every $t_0$. The remaining factor $\mc D(\bk) \lvert g_\bk \rvert^2$ is bounded for continuous reservoirs (see Sec. \ref{sec:pure_dephasing}). Therefore, in continuous reservoirs,
	\begin{equation}\label{eq:Gamma_vac_linear}
	\Gamma_L^{(vac)}=O(L).
	\end{equation}
	This says that in continuous reservoirs, vacuum fluctuations cannot cause superdecoherence.
	
	Now consider the excitation contribution $\xi^{(ex)}(t_0,\bk)=\mc D(\bk)\lvert g_\bk\rvert^2 \tau(t_0,\om_\bk)\,2\bar N_\bk$ \mbox{[Eq.~(\ref{eq:xi_ex})]}. It is bounded if both $\mc D(\bk)\lvert g_\bk\rvert^2$ and $\bar N_\bk$ are bounded. By Eq.~(\ref{eq:no_superdecoherence}) we have, in that case,
	\begin{equation*}
	\Gamma_L^{(ex)}=O(L).
	\end{equation*}
	Together with Eq.~(\ref{eq:Gamma_vac_linear}), this says there is no superdecoherence in continuous reservoirs.
	
	Conversely, we can consider the situations in which $\xi^{(vac/ex)}$ is unbounded. First, consider $\xi^{(vac)}$. It is unbounded if the reservoir is discrete, that is, if $\mc D(\bk)=\sum_{\bk'\in D}\delta(\bk-\bk')$. Even though, in this case, the conditions of Eq.~(\ref{eq:no_superdecoherence}) are not satisfied, this does not lead to superdecoherence per se. It is clear that $\Gamma^{(vac)}_L$ scales superlinearly with $L$ only when one of the modes in $D$ coincides exactly with a mode to which the matrix element is superlinearly susceptible. This is also illustrated by Fig.~\ref{fig:susceptibility} and Eq.~(\ref{eq:Gamma_sum}): there is superdecoherence in the array model when the state is $\GHZp$, and $\pi/a \in D$. If, in the array model, $\pi/a \notin D$, but instead $\pi/a+\delta\in D$, with $0<|\delta|\ll 1$, there is no superdecoherence. Note this in an asymptotic statement, and that, in the latter situation ($\pi/a \notin D$, $\pi/a+\delta\in D$), and for finite $L$, linear scaling of $\Gamma_L$ with $L$ only occurs after $1/L$ is approximately smaller than $|\delta|$. Further discussion on finite-size effects can be found in Sec.~\ref{sec:finite_size}.
	
	Secondly, consider $\xi^{(ex)}$. It is unbounded if the reservoir is discrete, like in the previous paragraph. It may additionally be unbounded if $\bar N_\bk$ is unbounded. This happens when a mode $\bk$ is excited but no modes in its neighborhood are excited. Again, this does not need to lead to superdecoherence per se. It is only when $\bk$ coincides exactly with a mode the matrix element is highly susceptible to that superlinear scaling of $\Gamma^{(ex)}_{\bi-\bj}$ is possible.

\subsection{Completely thermal reservoirs}

If the reservoir has a continuous spectrum, and the initial reservoir state is completely thermal, $\xi^{(ex)}(t_0,\bk)$ [Eq.~(\ref{eq:xi_ex})] is possibly unbounded because $\bar N_\om^{th}$ [Eq.~(\ref{eq:thermal_spectrum})] has an algebraic divergence at the origin. In this subsection, we show this nevertheless does not lead to superdecoherence (i.e. it does not lead to quadratic scaling of the decoherence function with the system size). However, superlinear scaling may be obtained, but only in subohmic reservoirs at nonzero temperature.

This is shown as follows. Consider $\xi^{(ex)}$  with $\bar N_\bk=\bar N_\om^{(th)}$. Note that $\tau$ [Eq.~(\ref{eq:tau})] is constant to first order at the origin, so that it cannot contribute to the divergence. Thus, $\xi^{(ex)}$ is bounded near the origin if $\mc D(\bk) \lvert g_\bk \rvert^2$ goes to zero fast enough near the origin. In the remainder of this subsection, we will assume the isotropic setting of Eq.~(\ref{eq:isotropic}), with $J(\om)$ as in Eq.~(\ref{eq:spectral_density}). Then the condition for bounded $\xi^{(ex)}$ becomes $d\geq 1$. This means there is no superlinear scaling of the decoherence function for Ohmic ($d=1$) and superohmic ($d>1$) continuous thermal reservoirs. 

\subsubsection{Subohmic thermal reservoirs}

For subohmic reservoirs ($d<1$), $\xi^{(ex)}$ in fact diverges at the origin. Here, we show how this can only lead to superlinear scaling of $\Gamma^{(ex)}_L$ with $L$ when $\gamma_L$ scales superlinearly with $L$ near the origin. Even if $\gamma_L$ scales superlinearly with $L$ near the origin, quadratic scaling may be approached, but not attained.

Let us first single out the divergence near the origin by defining $\Gamma^{(ex)}_L=\mc I_L + \mc J_L$, with
\begin{align}\label{eq:mathcalI_L}
\mc I_L& = \int_0^{\ep}\dd\om\, \gamma_L(\om)\, \xi^{(ex)}(t_0,\om),\\
		\xi^{(ex)}(t_0,\bk)&=J(\om)\tau(t_0,\om_\bk)\,2\bar N^{th}_\om,
\end{align}
and $\mc J_L$ the remainder of the integral. Note $\xi^{(ex)}(t_0,\om)$ now contains the thermal occupation density explicitly.

The integral $\mc J_L$ is $O(L)$ because, on the domain of integration, $\xi^{(ex)}$ is bounded [also see Eq.~(\ref{eq:no_superdecoherence})]. We now turn to $\mc I_L$. Given an $\ep$, there exists a constant $C_2$ such that $\xi^{(ex)} \leq C_2 \om^{d-1}$ on $(0,\ep]$. Thus,
\begin{equation*}
\mc I_L\leq C_2 \int_0^\ep \dd \om\, \gamma_L(\om) \om^{d-1}.
\end{equation*}
Since $\om^{d-1}$ is monotonically decreasing, the largest possible value of $\mc I_L$ occurs when $\gamma_L(\om)$ is peaked at low $\om$. Herein it is constrained by $\gamma_L(\om)\leq L^2$ [see Eq.  (\ref{eq:gamma_leq_L2})] and $\int_0^{2\pi/V}\dd \om\, \gamma_L(\om) \leq 2\pi L/V$ [Eq.~(\ref{eq:parseval})]. Under these constraints $\mc I_L$ is largest when $\gamma_L( \om)$ is a bump function, where the bump height is $L^2$, the left of the bump coincides with the origin, and the width of the bump is $2\pi/(VL)$. Therefore,
\begin{align}
	\mc I_L &\leq C_2 L^2 \int_0^{2\pi/(VL)}\dd \om \, \om^{d-1}\nn\\
	&=C_2 L^2\frac{1}{d}\left(\frac{2\pi}{V L}\right)^{d}\nn\\
	&=O\left(L^{2-d}\right).\label{eq:mc_I_result}
\end{align}
Thus, quadratic scaling of $\Gamma_L^{(ex)}(t_0)$, and thereby quadratic scaling of $\Gamma_L(t_0)$, cannot be obtained in subohmic continuous thermal reservoirs.

To approach superlinear scaling, it is essential that a superlinear peak of $\gamma_L(\om)$ must be able to approach the origin arbitrarily closely as a function of $L$. In fact, if, on the contrary, there is a $\delta>0$ such that $\gamma_L(\om)=O(L)$ for all $0\leq x\leq \delta$, then $\mc I_L=O(L)$. This is shown as follows. Assume there is a $\delta>0$ such that $0<\delta<\ep$ and $\gamma_L(\om)=O(L)$ for all $0\leq \om \leq \delta$. Then because $\om^{d-1}$ is finite on $[\delta,\ep]$, and because there is a constant $C_3$ such that $\gamma_L(\om)\leq C_3 L$ for all $[0, \delta)$, we have
\begin{align*}
	\mc I_L&\leq C_2 \int_0^{\delta}\dd \om\, \gamma_L(\om) \om^{d-1} + C_2\int_\delta^{\ep}\dd \om \, \gamma_L(\om) \om^{d-1}\\
	&\leq C_2 C_3 L \int_0^{\delta} \dd \om \, \om^{d-1} + O(L)\\
	&=O(L).
\end{align*}

An example in which this occurs is the array model, in the specific case that the dephasing susceptibility is given by $\gamma_{\mrm{GHZ}^{'}}(\om)$ [Eq.~(\ref{eq:closed_forms})]. To show this, let $\delta=\pi/(2a)$. Then $\cos^2( a\om/2)\geq\cos^{2}(\pi/4)\geq1/2$ for all $0\leq \om \leq\delta$, and thus $\gamma_{\mrm{GHZ}^{'}} \leq 2 \sin^2( a \om L/2)\leq 2$ for all $0\leq \om\leq\delta$. This means that the off-diagonal matrix element of $\GHZp$ does not suffer from superdecoherence in subohmic thermal reservoirs, despite the fact that $\xi^{(ex)}$ is unbounded at the origin.

The result Eq.~(\ref{eq:mc_I_result}) is an upper bound, so the question remains if it may be attained. This is not clear a priori because a dephasing susceptibility cannot attain the form of a bump function as in the proof. This is because it is the spectral density of a vector with a finite number of elements [Eq.~(\ref{eq:gamma_spectral_density})]. We now show by explicit construction that the upper bound may also be attained. This construction is in the subohmic version of the array model (Sec. \ref{sec:dephasing_susceptibility}), with dephasing susceptibility $\gamma_\mrm{GHZ}$ [Eq.~(\ref{eq:closed_forms})]. Roughly speaking, our strategy is to show that $\gamma_\mrm{GHZ}(\om)$ is a close enough approximation of the bump function. There are two main steps. The first is to show that for all $0\leq \om \leq 1/(2 a L^2)$, we have $\gamma_\mrm{GHZ}(\om)\geq L^2-1$, or equivalently,
\begin{equation}\label{eq:tilde_gamma_FM}
 \gamma'_\mrm{GHZ}(\om)\defeq 1-\frac{\gamma_\mrm{GHZ}(\om)}{L^2}\leq \frac{1}{L^2}.
\end{equation}
Consider the expansion of $\gamma'_\mrm{GHZ}(\om)$ in $a\om$ around $a\om=0$. Using the original definition of the dephasing susceptibility [Eq.~(\ref{eq:gamma})], we have $\gamma'_\mrm{GHZ}(\om)=\sum_{j=2,4,\ldots} c_j(a\om)^j$, with
\begin{equation*}
	|c_j|=\frac{1}{j!\,L^2}\sum_{mn}(m-n)^j<\frac{L^j}{j!}.
\end{equation*}
The radius of convergence of the expansion is infinite. Using the coefficients, we have
\begin{align*}
	\gamma'_\mrm{GHZ}(\om)&< \sum_{j=2,4,\ldots}\frac{L^j}{j!}\, (a\om)^j \nn \\
	&=\sum_{j=1,2,\ldots}\frac{1}{(2j)!}( a \om L)^{2j} \nn \\
	&\leq e^{( a\om L)^2}-1 \nn \\
	&\leq 4( a\om L)^2. \qquad \qquad (0\leq a\om L \leq 1)
\end{align*}
The last step can be checked most easily by plotting both functions. The last inequality holds specifically for $ a\om \leq 1 /(2L^2)$. After substitution, we have, therefore, that $\gamma'_\mrm{GHZ}(\om) \leq 1/L^2 $ for all $0\leq \om \leq 1/(2aL^2)$.

The second step is to show that  Eq.~(\ref{eq:tilde_gamma_FM}) enables us to approach quadratic scaling of $\mc I_L$ with $L$ arbitrary closely. First, note that, from Eq.~(\ref{eq:mathcalI_L}),
\begin{equation*}
\mc I_L > \int_0^{1/(2 a L^2)}\dd \om\, \gamma_\mrm{GHZ}(\om)\xi^{(ex)},
\end{equation*}
for $1/(2aL^2)<\ep$. There exists an $L_0$ and a constant $C_4$ such that for all $L>L_0$, $\xi^{(ex)}(t_0,\om)\geq C_4 \om^{d-1}$ on the entire domain of integration. Informally, this means that there is a $C_4$ such that, close enough to the origin, $\xi^{(ex)}(t_0,\om)\geq C_4 \om^{d-1}$. Thus, for this $C_4$,
\begin{equation*}
\mc I_L \geq C_4 \int_0^{1/(2aL^2)} \dd \om\, \gamma_\mrm{GHZ}(\om) \om^{d-1}.
\end{equation*}
Now using Eq.~(\ref{eq:tilde_gamma_FM}), this leads to
\begin{align*}
	\mc I_L &\geq C_4 (L^2-1) \frac{1}{d} \left(\frac{1}{2a L^2}\right)^{d}\\
	&=\Om\left[L^{2(1-d)}\right].
\end{align*}
Here the meaning of $\Om(x)$ is similar to that of $O(x)$, but $\Om(x)$  refers to a lower instead of an upper bound.\footnote{Formally, $f(x)=O[g(x)]$ means there exist an $x_0$ and a $c>0$ such that $|f(x)|\leq c g(x)$ for all $x>x_0$. The notation $\Om[f(x)]$ means there exist an $x_0$ and a $c>0$ such that $|f(x)|\geq c g(x)$ for all $x>x_0$.}

 Thus, in the array model with a subohmic continuous thermal reservoirs, quadratic scaling of $\Gamma_L^{(ex)}$, and thereby $\Gamma_L$, may be approached arbitrarily closely by the off-diagonal matrix element of $\GHZ$.

\subsection{Infinite time limit}\label{sec:infinite_time_limit} 
In our discussion of the system size scaling until now, we assumed the time $t$ to be fixed. Here we consider the infinite time limit of the isotropic case [Eq.~(\ref{eq:isotropic})], with $J(\om)$ as given in Eq.~(\ref{eq:spectral_density}). In the following, we no longer assume $d<1$ and $\bar N_\om=\bar N_\om^{th}$ as in the previous subsection. With $\partial_t \tau(t,\om)=\sin(\om t)/\om$ [cf. Eq.~(\ref{eq:tau})],
\begin{align}\label{eq:infinite_future}
	\lim_{t\to\infty} \partial_t \Gamma_L (t)
	=\frac{\pi}{2}\lim_{\om \downarrow 0} J(\om) \tilde \gamma(\om) (1+2\bar N_\om)	. 
\end{align}
Thus, the infinite time behavior of $\Gamma_L(t)$ depends only on the integrand at the origin, which is always nonnegative.
If the limit on the right hand side of Eq.~(\ref{eq:infinite_future}) is positive, $\Gamma_L(t)$ keeps growing indefinitely as a function of $t$. If, on the other hand, this limit is zero, $\Gamma_L(t)$ increases at most sublinearly with $t$ as $t$ goes to infinity. We call this a quasi-plateau, which naturally includes proper plateaus. These proper plateaus are also referred to as incomplete dephasing \cite{doll2007incomplete} or coherence trapping \cite{addis2014coherence}. In Appendix~\ref{sec:infinite_time_limit_appendix} we compute the height of the proper plateaus of $\Gamma_L^{(vac)}$ explicitly in the array model.

As an example, we can read off that for $\gamma_{\mrm{GHZ}^{'}}$, which is $O(\om^2)$ as $\om$ goes to zero [see Eq.~(\ref{eq:closed_forms})], in a completely thermal reservoir [Eq.~(\ref{eq:thermal_spectrum})], a (quasi-)plateau is reached for all $T\geq 0$ and $d\geq0$. From Eq.~(\ref{eq:infinite_future}) alone we cannot defer anything about the height of the \mbox{(quasi-)}plateau.

\section{Finite-size effects}\label{sec:finite_size}
In the previous section, we focused on the asymptotic system size scaling of the decoherence function. We saw that, in that case, a sharp delineation could be placed between cases of superlinear and linear scaling. For finite system sizes, the situation becomes less clear. This is because the decoherence function may scale quadratically up to some potentially large system size $L_0$, and show linear scaling only for $L>L_0$. Even though the main goal of this manuscript is to investigate the asymptotic scaling of the decoherence function with the system size, we discuss some finite-size effects in this section.

\subsection{Time}
Assume, for simplicity, a linear, isotropic dispersion relation, $\om_\bk = \om = |\bk|$, in units where the proportionality constant equals unity. Consider the temporal factor $\tau(t,\om_\bk)$ [Eq.~(\ref{eq:tau})] as a function of $\bk$. The function is peaked at the origin, with height $t^2/2$. Away from the origin, it drops to zero at $|\bk|=2\pi/t$ and remains small afterwards [$O(1/|\bk|^2)$]. Thus, for large $t$, $\tau$ gives large weight to wave vectors with a length below $2\pi/t$, and ever smaller weight to wave vectors with a length above $2\pi/t$.

In Sec. \ref{sec:dephasing_susceptibility}, we showed that, if $\gamma_L(\bk)$, as a function of $\bk$, has a peak of height $L^2$, the support of that peak must scale as $1/L$. This effect causes the absence of superdecoherence in continuous reservoirs. However, if this peak is located at the origin, but $t$ is such that the peak of $\tau(t,\om_\bk)$ is much narrower than that of $\gamma_L(\bk)$, we have that $\gamma_L(\bk)$ is approximately constant on the interval where $\tau(t,\om_\bk)$ is non-negligible. Thus, the reducing bandwidth of $\gamma_L(\bk)$ is only guaranteed to have an effect if
\begin{equation}\label{eq:short_time_condition}
a L \gtrsim t.
\end{equation}
Therefore, the actual scaling of $\Gamma_L$ as a function of $L$ may approach its asymptotic scaling only at times small compared to the system size.

\begin{figure}
	
	\hbox{ \hspace{-1em}
		
		\begin{tikzpicture}
		\begin{axis}[cycle list name=exotic,axis lines = left,ymax=1000, ylabel=$\Gamma_\mrm{GHZ}$, ytick={0,400,800}, xlabel=$L$, width=.55\columnwidth,
		y label style={at={(axis description cs:.4,1.1)},rotate=-90,anchor=south},
		legend style={at={(0.04,.8)},anchor=west}
		]
		\addplot+[mark options={scale=.4}]
		coordinates {
			(0,0.0)(1,5.991467672098216)(2,17.969406177938684)(3,34.54184616979266)(4,54.8854980932753)(5,78.40705466223449)(6,104.63656899281624)(7,133.17965703716467)(8,163.69166909883137)(9,195.86188045022996)(10,229.4028147028151)(11,264.0423501331116)(12,299.51730287127634)(13,335.56762314495455)(14,371.93047427883545)(15,408.3333537946399)(16,444.4849587890211)(17,480.061268956324)(18,514.6808137690218)(19,547.8506539582286)(20,578.7964201719748)(21,604.4438635504418)(22,627.717077634952)(23,649.2393315573134)(24,669.3505442719347)(25,688.2762077451877)(26,706.1802665857003)(27,723.1885648294227)(28,739.4012697379726)(29,754.9002250633918)(30,769.7536465841149)
		};
		\addlegendentry{\scriptsize $d=1$}
		\addplot+[mark options={scale=.45}]
		coordinates {
			(0,0.0)(1,19.999875000781245)(2,40.09924874376585)(3,60.223349246461545)(4,80.35829024830996)(5,100.49919815797698)(6,120.64380344825085)(7,140.79085726040418)(8,160.939587344284)(9,181.08946887563005)(10,201.2401122693162)(11,221.39120009510142)(12,241.54244505619246)(13,261.6935544021824)(14,281.84418918571447)(15,301.99390212187393)(16,322.1420188812765)(17,342.28736316989676)(18,362.42746294056803)(19,382.5553445324745)(20,402.63359494599666)(21,402.71206410800494)(22,402.74085497022105)(23,402.75733190592945)(24,402.7684168222794)(25,402.7765250107791)(26,402.7827687072913)(27,402.78774791034357)(28,402.7918212963821)(29,402.7952193121851)(30,402.798098144165)
		};
		\addlegendentry{\scriptsize $d=2$}
		\end{axis}		
		\end{tikzpicture}\hspace{1em}\begin{tikzpicture}
		\begin{axis}[cycle list name=exotic,axis lines = left,ymax=1000, ytick={0,400,800}, ylabel=$\Gamma_{\mrm{GHZ}^{'}}$, xlabel=$L$, width=.55\columnwidth,
		y label style={at={(axis description cs:.4,1.1)},rotate=-90,anchor=south},
		legend style={at={(0.04,.8)},anchor=west}
		]
			\addplot+[mark options={scale=.4}]
		coordinates {
			(0,0.0)(1,19.999875000781245)(2,39.90025125935913)(3,59.825354277648096)(4,79.73961679678435)(5,99.6598462237392)(6,119.57637827008716)(7,139.49535883831456)(8,159.41266313481543)(9,179.3311188787825)(10,199.24881276040958)(11,219.1669510741357)(12,239.08493225255597)(13,259.0027778158751)(14,278.92109794165225)(15,298.8384962200567)(16,318.7574906752183)(17,338.67371265959764)(18,358.5951791619259)(19,378.50442748548954)(20,398.46330698743776)(21,398.4224052378723)(22,398.4311817880989)(23,398.4276444118178)(24,398.429499054895)(25,398.428376970122)(26,398.42911937733624)(27,398.428597291091)(28,398.42898102186007)(29,398.42868938239366)(30,398.42891692675045)
		};
		\addlegendentry{\scriptsize $d=2$}
		\addplot+[mark options={scale=.4}]
		coordinates {
			(0,0.0)(1,5.991467672098216)(2,5.996464510454182)(3,10.59596283482366)(4,11.424249227564484)(5,15.430440265781847)(6,16.72867354237664)(7,20.340480532738137)(8,21.983363505781355)(9,25.28444576855644)(10,27.21480513014498)(11,30.243765669444834)(12,32.43730890087632)(13,35.20621966782114)(14,37.66259957456313)(15,40.15900786322851)(16,42.90669067331687)(17,45.07907865632672)(18,48.20823199394156)(19,49.887680708065396)(20,53.79120339764993)(21,52.39640325195532)(22,53.375832400217675)(23,52.60430138633153)(24,53.24381158018571)(25,52.697772532671706)(26,53.17333811789793)(27,52.75314310633386)(28,53.12854142994254)(29,52.79019017042111)(30,53.097372715596265)
		};
		\addlegendentry{\scriptsize $d=1$}
		\end{axis}		
		\end{tikzpicture}
		
	}
	\caption{
		\label{fig:time}	
		The decoherence function in the array model (see Sec.~\ref{sec:dephasing_susceptibility}), as a function of $L$, for the off-diagonal matrix element of the state $\GHZ$ (left) and $\GHZp$ (right). In both plots, we use units where $a=1$, and set $t=20$, $J(\om)=\alpha_d \om^d e^{-\om/\om_c}$,  with $\alpha_d=1$ (for both $d=1$ and $d=2$), $\om_c=20$, and $\bar N_\om=0$. We have used the analytical expressions for the decoherence function that are derived in appendix \ref{sec:explicit_expressions_for_the_vacuum_contribution}. (\textbf{Left}) For $d=1$ the decoherence function increases quadratically initially, after which it scales (sub)linearly. For $d=2$ there is no quadratic scaling, even for $aL\ll t$. (\textbf{Right}) No superlinear scaling for any $t$, $L$ and $d$ (including $d$ other than $d=1,2$, which are not shown). The lines for $d=2$ in the left and right plot are similar, but not exactly equal.  	}
\end{figure}
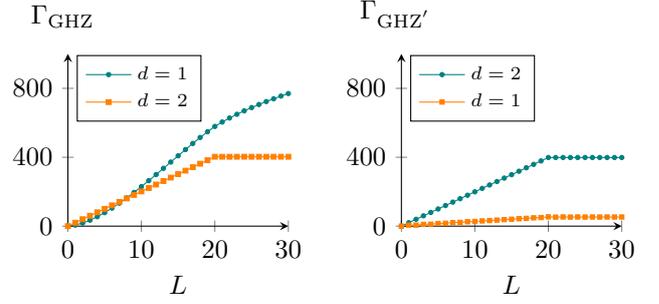

This seems to form an important caveat to our asymptotic results. However, it only applies in special cases. Firstly, $\gamma_L(\om)$ needs to scale superlinearly as a function of $L$ near the origin, which is rarely the case [see App. \ref{sec:typical_dephasing_susceptibility}]. Secondly, even if $\gamma_L(\om)$ scales superlinearly, the remaining factors $\xi^{(ex)}$ and $\xi^{(vac)}$ may kill the entire integrand around the origin [see Eq.~\mbox{(\ref{eq:Gamma_vac_ex})}], for example when $\xi^{(vac)}(t,\om)=O(\om^d)$ and $\xi^{(ex)}=O(\om^d)$ as $\om\to 0$, with $d\geq1$. Then for every $\gamma_L(\om)$ that scales superlinearly at the origin and fixed $L_0$, there is a continuous crossover from superlinear to linear behavior in $L$ around $L_0$ as a function of $d$. See Fig.~\ref{fig:time} for two concrete examples.

\subsection{Peaked occupation density}
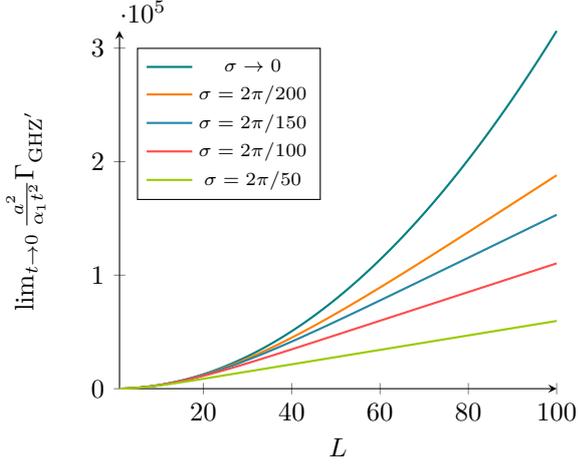
\begin{figure}
	\begin{center}
	\begin{tikzpicture}
	\begin{axis}[cycle list name=exotic,axis lines = left, ylabel=$\lim_{t\rightarrow  0} \frac{a^2}{\alpha_1 t^2} \Gamma_{\mrm{GHZ}^{'}}$, xlabel=$L$, width=.9\columnwidth,
	legend style={at={(0.04,.74)},anchor=west}
	]
	\addplot+[mark=none, thick]
	coordinates {
		(1,41.2855)(2,145.403)(3,312.352)(4,542.133)(5,834.746)(6,1190.19)(7,1608.47)(8,2089.58)(9,2633.52)(10,3240.29)(11,3909.89)(12,4642.33)(13,5437.60)(14,6295.70)(15,7216.63)(16,8200.39)(17,9246.99)(18,10356.4)(19,11528.7)(20,12763.8)(21,14061.7)(22,15422.4)(23,16846.0)(24,18332.4)(25,19881.7)(26,21493.8)(27,23168.7)(28,24906.4)(29,26707.0)(30,28570.4)(31,30496.7)(32,32485.7)(33,34537.6)(34,36652.4)(35,38829.9)(36,41070.3)(37,43373.6)(38,45739.6)(39,48168.5)(40,50660.3)(41,53214.8)(42,55832.2)(43,58512.4)(44,61255.5)(45,64061.4)(46,66930.1)(47,69861.7)(48,72856.0)(49,75913.3)(50,79033.3)(51,82216.2)(52,85461.9)(53,88770.4)(54,92141.8)(55,95576.0)(56,99073.0)(57,102633.)(58,106256.)(59,109941.)(60,113690.)(61,117501.)(62,121375.)(63,125312.)(64,129311.)(65,133374.)(66,137499.)(67,141687.)(68,145938.)(69,150252.)(70,154629.)(71,159068.)(72,163571.)(73,168136.)(74,172764.)(75,177455.)(76,182208.)(77,187025.)(78,191904.)(79,196846.)(80,201851.)(81,206919.)(82,212050.)(83,217243.)(84,222500.)(85,227819.)(86,233201.)(87,238646.)(88,244153.)(89,249724.)(90,255357.)(91,261053.)(92,266812.)(93,272634.)(94,278519.)(95,284466.)(96,290477.)(97,296550.)(98,302686.)(99,308885.)(100,315146.)
	};
	\addlegendentry{\scriptsize $\sigma\rightarrow 0$}
		\addplot+[mark=none, thick]
	coordinates {
		(1,41.2855)(2,145.372)(3,312.166)(4,541.514)(5,833.199)(6,1186.95)(7,1602.42)(8,2079.22)(9,2616.90)(10,3214.96)(11,3872.81)(12,4589.86)(13,5365.43)(14,6198.80)(15,7089.22)(16,8035.86)(17,9037.88)(18,10094.4)(19,11204.4)(20,12367.0)(21,13581.2)(22,14846.0)(23,16160.2)(24,17522.8)(25,18932.8)(26,20388.8)(27,21889.9)(28,23434.8)(29,25022.4)(30,26651.5)(31,28320.9)(32,30029.4)(33,31775.8)(34,33558.9)(35,35377.5)(36,37230.4)(37,39116.5)(38,41034.6)(39,42983.5)(40,44962.0)(41,46969.1)(42,49003.6)(43,51064.4)(44,53150.4)(45,55260.6)(46,57393.9)(47,59549.4)(48,61725.9)(49,63922.6)(50,66138.6)(51,68372.8)(52,70624.4)(53,72892.6)(54,75176.5)(55,77475.3)(56,79788.2)(57,82114.5)(58,84453.4)(59,86804.3)(60,89166.4)(61,91539.2)(62,93922.0)(63,96314.2)(64,98715.3)(65,101125.)(66,103542.)(67,105966.)(68,108398.)(69,110836.)(70,113279.)(71,115729.)(72,118183.)(73,120643.)(74,123107.)(75,125575.)(76,128047.)(77,130523.)(78,133002.)(79,135485.)(80,137970.)(81,140458.)(82,142948.)(83,145440.)(84,147935.)(85,150432.)(86,152930.)(87,155430.)(88,157932.)(89,160435.)(90,162939.)(91,165444.)(92,167950.)(93,170458.)(94,172966.)(95,175475.)(96,177985.)(97,180495.)(98,183006.)(99,185518.)(100,188030.)
	};
	\addlegendentry{\scriptsize $\sigma=2\pi/200$}
		\addplot+[mark =none, thick]
	coordinates {
		(1,41.2855)(2,145.348)(3,312.022)(4,541.034)(5,832.001)(6,1184.44)(7,1597.75)(8,2071.26)(9,2604.16)(10,3195.59)(11,3844.57)(12,4550.05)(13,5310.91)(14,6125.95)(15,6993.89)(16,7913.40)(17,8883.11)(18,9901.58)(19,10967.3)(20,12078.9)(21,13234.6)(22,14433.1)(23,15672.6)(24,16951.6)(25,18268.6)(26,19621.9)(27,21009.8)(28,22431.0)(29,23883.7)(30,25366.4)(31,26877.7)(32,28416.0)(33,29979.9)(34,31568.0)(35,33178.9)(36,34811.2)(37,36463.7)(38,38135.1)(39,39824.1)(40,41529.8)(41,43250.8)(42,44986.3)(43,46735.1)(44,48496.3)(45,50269.0)(46,52052.4)(47,53845.6)(48,55647.8)(49,57458.3)(50,59276.5)(51,61101.7)(52,62933.3)(53,64770.8)(54,66613.6)(55,68461.2)(56,70313.3)(57,72169.5)(58,74029.2)(59,75892.2)(60,77758.2)(61,79626.9)(62,81498.0)(63,83371.2)(64,85246.3)(65,87123.2)(66,89001.6)(67,90881.4)(68,92762.5)(69,94644.6)(70,96527.7)(71,98411.6)(72,100296.)(73,102182.)(74,104068.)(75,105954.)(76,107841.)(77,109728.)(78,111616.)(79,113504.)(80,115392.)(81,117280.)(82,119169.)(83,121058.)(84,122947.)(85,124836.)(86,126725.)(87,128614.)(88,130504.)(89,132393.)(90,134283.)(91,136172.)(92,138062.)(93,139952.)(94,141841.)(95,143731.)(96,145621.)(97,147510.)(98,149400.)(99,151290.)(100,153180.)
	};
	\addlegendentry{\scriptsize $\si=2\pi/150$}
		\addplot+[mark =none, thick]
	coordinates {
		(1,41.2855)(2,145.279)(3,311.610)(4,539.667)(5,828.602)(6,1177.34)(7,1584.61)(8,2048.91)(9,2568.59)(10,3141.81)(11,3766.62)(12,4440.90)(13,5162.47)(14,5929.05)(15,6738.31)(16,7587.86)(17,8475.32)(18,9398.30)(19,10354.4)(20,11341.4)(21,12356.8)(22,13398.6)(23,14464.5)(24,15552.6)(25,16660.8)(26,17787.3)(27,18930.4)(28,20088.3)(29,21259.6)(30,22442.9)(31,23636.8)(32,24840.1)(33,26051.8)(34,27270.8)(35,28496.2)(36,29727.2)(37,30963.0)(38,32203.1)(39,33446.8)(40,34693.6)(41,35943.1)(42,37194.9)(43,38448.6)(44,39703.9)(45,40960.6)(46,42218.5)(47,43477.3)(48,44736.9)(49,45997.2)(50,47258.1)(51,48519.4)(52,49781.0)(53,51043.0)(54,52305.2)(55,53567.6)(56,54830.2)(57,56092.9)(58,57355.7)(59,58618.6)(60,59881.5)(61,61144.5)(62,62407.6)(63,63670.7)(64,64933.8)(65,66196.9)(66,67460.0)(67,68723.2)(68,69986.3)(69,71249.5)(70,72512.6)(71,73775.8)(72,75039.0)(73,76302.2)(74,77565.4)(75,78828.5)(76,80091.7)(77,81354.9)(78,82618.1)(79,83881.3)(80,85144.4)(81,86407.6)(82,87670.8)(83,88934.0)(84,90197.2)(85,91460.4)(86,92723.5)(87,93986.7)(88,95249.9)(89,96513.1)(90,97776.3)(91,99039.5)(92,100303.)(93,101566.)(94,102829.)(95,104092.)(96,105355.)(97,106619.)(98,107882.)(99,109145.)(100,110408.)
	};
	\addlegendentry{\scriptsize $\si=2\pi/100$}
	\addplot+[mark=none, thick]
	coordinates {
		(1,41.2855)(2,144.909)(3,309.410)(4,532.434)(5,810.833)(6,1140.81)(7,1518.07)(8,1938.00)(9,2395.85)(10,2886.83)(11,3406.35)(12,3950.04)(13,4513.88)(14,5094.26)(15,5688.02)(16,6292.40)(17,6905.12)(18,7524.24)(19,8148.23)(20,8775.86)(21,9406.16)(22,10038.4)(23,10672.0)(24,11306.6)(25,11941.8)(26,12577.5)(27,13213.5)(28,13849.6)(29,14486.0)(30,15122.4)(31,15758.8)(32,16395.3)(33,17031.8)(34,17668.3)(35,18304.8)(36,18941.3)(37,19577.9)(38,20214.4)(39,20850.9)(40,21487.4)(41,22124.0)(42,22760.5)(43,23397.0)(44,24033.5)(45,24670.1)(46,25306.6)(47,25943.1)(48,26579.6)(49,27216.2)(50,27852.7)(51,28489.2)(52,29125.8)(53,29762.3)(54,30398.8)(55,31035.3)(56,31671.9)(57,32308.4)(58,32944.9)(59,33581.4)(60,34218.0)(61,34854.5)(62,35491.0)(63,36127.5)(64,36764.1)(65,37400.6)(66,38037.1)(67,38673.7)(68,39310.2)(69,39946.7)(70,40583.2)(71,41219.8)(72,41856.3)(73,42492.8)(74,43129.3)(75,43765.9)(76,44402.4)(77,45038.9)(78,45675.4)(79,46312.0)(80,46948.5)(81,47585.0)(82,48221.6)(83,48858.1)(84,49494.6)(85,50131.1)(86,50767.7)(87,51404.2)(88,52040.7)(89,52677.2)(90,53313.8)(91,53950.3)(92,54586.8)(93,55223.3)(94,55859.9)(95,56496.4)(96,57132.9)(97,57769.5)(98,58406.0)(99,59042.5)(100,59679.0)
	};
	\addlegendentry{\scriptsize $\si=2\pi/50$}
	\end{axis}		
	\end{tikzpicture}
	\end{center}
	\vspace{-1em}
	\caption{
		\label{fig:finite_size}	The leading order in time of the decoherence function of the off-diagonal matrix element of $\GHZp$, as a function of the system size $L$. The setting is that of the array model (Sec.~\ref{sec:dephasing_susceptibility}), with $d=1$. In units where $a=1$, the occupation density $\bar N_\om$ is taken to be a Gaussian, with mean $\om_0=\pi$, standard deviation $\si$ and integrated occupation $\bar N_{tot}=10$. We see linear scaling with $L$ is obtained for $1 / L \lesssim \si$. In the limit $\si \rightarrow 0$, the occupation density becomes unbounded, and consequently, it is only in the limit that  the decoherence function scales as $L^2$ for all $L$. The plot uses an analytic solution of Eq. (\ref{eq:isotropic}), with $J(\om)=\frac{\al_1}{2}\om [ 1-\Theta(\om-2\pi)]$, where $\Theta$ is the step function. This form of the spectral density is chosen to accentuate the finite-size effects; the function $\gamma_{\mrm{GHZ}^{'}}(\om)$ has peaks at $\om=\pi,\pi+2\pi,\ldots$, whereas, in this example, $\bar N_\om$ only has a peak at $\om_0=\pi$. Finite-size effects only occur at places where the two peaks overlap, and including frequencies higher than $2\pi$ into $J(\om)$ means including more effects that scale with $L$ rather than $L^2$.
	}
\end{figure}

A similar finite-size effect occurs if the occupation density has a peak that coincides with a superlinear peak of the dephasing susceptibility. To separate this effect from the one in the previous subsection, consider as an example the state $\GHZp$, in the array model,  with a Gaussian occupation density $\bar N_\om$. The Gaussian has mean $\pi/a$, variance $\si$ and an integrated number of bosons $\bar N_{tot}\defeq \int_{-\infty}^\infty \dd \om\, \bar N_\om$. That is, $\bar N_\om=\bar N_{tot}{\exp[-(\om-\pi/a)^2/(2\si^2)]}/(\sqrt{2\pi}\si)$. Similar to in the previous subsection, the mitigating effect of the $1/L$ bandwidth of the dephasing susceptibility has an effect only after the peak of the dephasing susceptibility becomes narrower than that of the occupation density. That is, we only expect linear scaling of the decoherence function for
\begin{equation*}
 \frac{2\pi}{aL} < \si. 
\end{equation*}
See Fig.~\ref{fig:finite_size} for plots of the leading order in time of the decoherence function $\Gamma_\mrm{GHZ^{'}}$.

Again, in many situations, the effect discussed in this subsection does not have significant effects.  Firstly, note that the integral in Eq.~(\ref{eq:isotropic}) is over infinitely many periods of the dephasing susceptibility. In the example above, the peak of the occupation density occurs only at a single frequency. In this case, the effect described in this subsection will thus only occur at one of the periods of $\gamma_{\mrm{GHZ}^{'}}(\om)$. Secondly, the center of the peak of the occupation density has to coincide exactly with the peak of the dephasing susceptibility. 

The latter situation, where there is a single peak in the occupation density that overlaps exactly with the peak in the dephasing susceptibility, occurred in the ion-trap experiment by Monz et al., where the model of single-reservoir dephasing is applicable \cite{monz201114}. The state $\GHZ$ was prepared in a semi-static, semi-uniform magnetic field, which was produced with a Helmholtz coil. Fluctuations of the field, caused by current fluctuations in the coil, excited long-wavelength modes (with $k\approx 0$). The dephasing susceptibility of the off-diagonal matrix element of $\GHZ$ scales as $L^2$ at $k=0$ (see Fig.~\ref{fig:susceptibility}), and exactly the modes $k \approx 0$ where heavily excited in the experiment. If modes were excited away from the origin, there would not have been superdecoherence. This follows from the explicit form of the dephasing susceptibility of the off-diagonal matrix element of $\GHZ$ (also see Fig.~\ref{fig:susceptibility}). Furthermore, even if modes near the origin were excited, almost any matrix element other than the off-diagonal element of $\GHZ$ would not have suffered superdecoherence. This statement is shown in more detail in Appendix~\ref{sec:typical_dephasing_susceptibility}.

\section{Conclusion and outlook}\label{sec:conclusion_and_outlook}

In this paper, we studied superdecoherence in the model of single-reservoir dephasing for asymptotic system sizes. We have shown that if the density of modes in $k$-space, $\mc D(\bk)$, and the occupation density $\bar N_\bk$ are bounded, superdecoherence is not possible. This is because if there is a $\bk$ such that the dephasing susceptibility scales quadratically with the system size, $\gamma_L(\bk)\propto L^2$, the support of this peak in the dephasing susceptibility always scales inversely with $L$.

Superdecoherence may thus only be obtained if $\mc D(\bk)$ or  $\bar N_\bk$ is unbounded. The former happens if the reservoir supports only a discrete set of modes. The latter happens if the reservoir supports a continuum of modes, but only perfectly isolated modes are excited. In both cases, the unbounded point must coincide exactly with the mode for which the dephasing susceptibility $\gamma_L(\bk)$ scales quadratically.

For completely thermal continuous reservoirs, the occupation density $\bar N_\om^{th}$ diverges algebraically at the origin. Nevertheless, even in this case, the decoherence function scales at most linearly with the system size. There is one exception. This is the subohmic continuous thermal reservoir with nonzero temperature, where, furthermore, the dephasing susceptibility must scale superlinearly (which includes quadratic scaling) at low frequencies. In this case, the decoherence function may approach, but not attain, quadratic scaling with the system size.

All effects discussed int this manuscript can in principle be observed experimentally. One could compare the effects of narrow-band versus broadband noise at a frequency to which the system is highly susceptible. Or, the system could be placed in a high-Q cavity, in the vacuum state, that supports exactly the mode the system is highly susceptible to. This is to be compared to a situation where the cavity is slightly longer.

Other applications lie in quantum metrology, in which superdecoherence can be used as a means of enhancing sensitivity. In this context, it is well-known that the GHZ state is highly susceptible to long-wavelength modes \cite{giovannetti2006quantum, giovanetti2004quantum}. The dephasing susceptibility, as defined in this paper, offers an effective way to extend metrology to other states and wavelengths; any state for which there is a $\om_0$ such that $\gamma_L(\om_0)\propto L^2$ is suitable for quantum metrology of the mode with wavelength $\om_0$. An example is a linear array of qubits, with spacing $a$, in the state $\GHZp$ [Eq.~(\ref{eq:GHZ})]. This system is highly susceptible to the staggered mode $\om_0=\pi/a$. The dephasing susceptibility shows the added benefit that with increasing system size, the array becomes less sensitive to frequencies other than $\om_0$.
\ \\

\textbf{Acknowledgments.} The authors would like to thank T. Bannink for discussion and L. Viola for comments on our first manuscript.


\clearpage

\appendix

\appendix
\appendixpage

\section{Spin-boson dephasing for arbitrary reservoir states}\label{sec:general_reservoir_state}
In this section, we generalize the decoherence function of single-reservoir dephasing, as it is found in Refs.~\mbox{\cite{palma1996quantum, reina2002decoherence,breuer2002theory}}, to more general reservoir states. We pay specific attention to Gaussian states, Gaussian product states, and a product of displaced thermal states. The latter solution is included in the main text as Eqs.~(\ref{eq:Gamma_single_qubit}) and~(\ref{eq:Gamma_sum}).  

For a completely general reservoir state, it can be shown that \cite{palma1996quantum, reina2002decoherence,breuer2002theory}
\begin{equation}\label{eq:structure}
	|\rho_{\bi\bj}(t)|=\left|\tilde \chi\left({\la}\right)\right|\, |\rho_{\bi\bj}(0)|,
\end{equation}
where $\tilde \chi$ is the \emph{characteristic function} of the reservoir state,
\begin{equation}\label{eq:characteristic_function_general}
\tilde \chi\left(\la\right) \defeq \left\langle e^{ \sum_{\bk\in D}(\la_\bk a^\dagger - \la_\bk^* a_\bk^{\nodagger})} \right\rangle_{\rho_B(0)}.
\end{equation}
Equations (\ref{eq:structure}) and (\ref{eq:characteristic_function_general}) hold for the interaction as well as the Schr\"odinger picture density operators. The set $D$ contains all wave vectors that are supported by the reservoir. 

The argument of the characteristic function, $\la\in \mathbb C^{|D|}$, depends on the matrix index $(\bi,\bj)$ and the time $t$, but the notation of this dependence is suppressed. The $\bk$th entry of $\la$ is given by \cite{palma1996quantum, reina2002decoherence,breuer2002theory}
\begin{equation}\label{eq:lambda}
\la_\bk = g_{\bk}\,  \tilde{\mrm{d}}^*(\bk) \,\frac{1-e^{i\om_\bk t}}{\om_\bk},
\end{equation}
with $\tilde{\mrm{d}}$ the Fourier transform of $\bd=\bi-\bj$ [see Eq.~(\ref{eq:spectral_density})]. The exponent in Eq.~(\ref{eq:lambda}) stems from the internal time evolution of the reservoir. Equations (\ref{eq:structure}) and (\ref{eq:characteristic_function_general}) give the most general form of the absolute value of the time evolved reduced density matrix in the single-reservoir dephasing model. 

For the class of \emph{Gaussian states} \cite{ferraro2005gaussian,adesso2014continuous}, the absolute value of the characteristic function is given by 
\begin{equation*}
	\lvert\tilde \chi (\la)\rvert=e^{-\Gamma(\la)},
\end{equation*}
with 
\begin{equation}\label{eq:Gamma_lambda_general}
\Gamma(\la)=\half \La^T \si \La.
\end{equation}
This is the most general form of the decoherence function. Here, writing $\la_{\bk_i}$ as $\la_i$ for short, 
\begin{equation*}
	\La^T=\sqrt{2}\,\left(\Re\,\la_1, \Im \, \la_1,\ldots, \Re\,\la_{|D|},\Im\,\la_{|D|}\right).
\end{equation*} 
The $2|D|\times2|D|$ matrix $\si$ is the \emph{covariance matrix},
\begin{equation} \label{eq:covariance_matrix}
	\si_{mn} =\shalf  \langle \{\hat R_m,\hat R_n\} \rangle-\langle \hat R_m \rangle \langle \hat R_n \rangle.
\end{equation}
The expectation value is with respect to the Gaussian initial reservoir state $\rho_B(0)$. The vector $\hat R$ is defined by
\begin{equation*}
	\hat R^T=(\hat q_1,\hat p_1,\ldots \hat q_{|D|},\hat p_{|D|}),
\end{equation*}
with $\{\cdot,\cdot\}$ the anti-commutator. To avoid confusion about operators versus numbers, in this section we write operators (and vectors containing operators) with hats, as opposed to in the main text. The quadrature operators $\hat q_m$ and $\hat p_m$, in turn, are defined by
\begin{equation*}
	\hat q_m = \frac{1}{\sqrt{2}}(\hat a_m + a^\dagger_m ), \qquad \hat p_m = \frac{1}{i\sqrt{2}}(\hat a_m^\nodagger - \hat a^\dagger_m).
\end{equation*}  
To obtain the decoherence function as a function of time and the density matrix index $(\bi,\bj)$, the expression for $\la$ [Eq.~(\ref{eq:lambda})] has to be inserted into Eq.~(\ref{eq:Gamma_lambda_general}).

We may consider various simplifications of the decoherence function as it is given in Eq.~(\ref{eq:Gamma_lambda_general}). If the reservoir modes are unentangled, that is, $\rho_B(0)=\bigotimes_\bk \rho_{B,\bk}(0)$ with all $\rho_{B,\bk}(0)$ Gaussian, the covariance matrix is block-diagonal. Each block corresponds to a $2\times 2$ single-mode covariance matrix, which we denote by $\si_\bk$. In this case, we may write
\begin{equation}\label{eq:Gamma_single_sum}
\Gamma(\la)= \sum_\bk \mathlarger{(}\Re \la_\bk,  \Im \la_\bk \mathlarger{)}
\, \si_\bk
\left( \begin{array}{c}
\Re \la_\bk\\  
\Im \la_\bk
\end{array}\right),
\end{equation}
where,  by Eq.~(\ref{eq:covariance_matrix}), the entries of the single-mode covariance matrix read
\begin{align*}
	(\si_\bk)_{11}&=\langle \hat q^2_\bk \rangle-\langle \hat q_\bk \rangle^2\\
	(\si_\bk)_{22}&=\langle \hat p^2_\bk \rangle-\langle \hat p_\bk \rangle^2 \\
	(\si_\bk)_{12}&=
	\half \langle \{\hat q_\bk,\hat p_\bk\}\rangle-\langle \hat q_\bk\rangle \langle \hat p_\bk\rangle \\
	(\si_\bk)_{21}&=(\si_\bk)_{12}.
\end{align*}

If the mode $\bk$ is initially in the thermal state, with temperature $T_\bk$, its density matrix reads $\rho_{B,\bk}(0)\propto e^{-\om_\bk \hat a_\bk^\dagger \hat a_\bk^\nodagger/T_{\bk}}$. In this case, the single-mode covariance matrix is diagonal,
\begin{equation} \label{eq:covariance_single_mode}
	\sigma_\bk = \mrm{diag}\left(\bar N_\bk+\shalf,\, \bar N_\bk +\shalf\right),
\end{equation}
with $\bar N_\bk$ the occupation number [Eq.~(\ref{eq:Bose-Einstein})]. A special thermal state is the vacuum, where $\bar N_\bk=0$.

If the reservoir modes are unentangled, and every mode is thermally excited with its own temperature, we have from combining Eqs.~(\ref{eq:Gamma_single_sum}) and~(\ref{eq:covariance_single_mode}) that 
\begin{equation}\label{eq:Gamma_lambda}
\Gamma(\la) =  \sum_\bk |\la_\bk|^2 (\bar N_\bk+\shalf).
\end{equation}
Inserting the equation for $\la_\bk$ [Eq.~(\ref{eq:lambda})], we obtain Eq.~(\ref{eq:Gamma_sum}).

In general, a single-mode Gaussian state can also be represented as a squeezed and displaced thermal state \cite{ferraro2005gaussian}, 
\begin{align*}
	(\si_\bk)_{11}&=(\bar N_\bk +\shalf)[\cosh(2r)+\sinh(2r)\cos(\varphi)]\\
	(\si_\bk)_{22}&=(\bar N_\bk +\shalf)[\cosh(2r)-\sinh(2r)\cos(\varphi)]\\
	(\si_\bk)_{12}&=
	-(\bar N_\bk +\shalf)\sinh(2r)\sin(\varphi)\\
	(\si_\bk)_{21}&=(\si_\bk)_{12}.
\end{align*}
Here $r$ is the squeezing magnitude, and $\varphi$ the squeezing angle. Note these expressions are invariant under displacement. Therefore, the decoherence function of a displaced thermal state is equal to Eq.~(\ref{eq:Gamma_lambda}), with $\bar N$ the regular Bose-Einstein distribution. Squeezing, on the other hand, does affect the covariance matrix, and would alter Eq.~(\ref{eq:Gamma_lambda}) straightforwardly. In the main text, we assume for simplicity that the reservoir modes are not squeezed. 

Displaced vacuum states are precisely the coherent states. Thus, even if a reservoir mode is in a highly excited coherent state, this mode does not contribute more to the dephasing process than the same mode in the vacuum state would have done. A mixture of coherent states does lead to extra dephasing. However, the only mixture that can be described in the Gaussian state formalism is the thermal state. 

To summarize, in single-reservoir dephasing, the decoherence process of the system is completely determined by the reservoir characteristic function; $\left|\tilde \chi\left({\la}\right)\right|= |\rho_{\bi\bj}(t)|/|\rho_{\bi\bj}(0)|$. The argument of the characteristic function, $\la$, is a complex vector which depends on the matrix index $(\bi,\bj)$ and time [Eq.~(\ref{eq:lambda})]. For completely general reservoir states, the characteristic function is given by Eq.  (\ref{eq:characteristic_function_general}).  For general Gaussian reservoir states, $|\tilde \chi(\la)|=e^{-\Gamma(\la)}$, with $\Gamma(\la)=\half \La^T \si \La$ the decoherence function, generalized to Gaussian states [Eq.~(\ref{eq:Gamma_lambda_general})]. In case the reservoir modes are unentangled, $\Gamma(\la)$ may be written using a single sum over $\bk$ [Eq.~(\ref{eq:Gamma_single_sum})]. If, furthermore, each of these modes is a (possibly) displaced thermal state, the decoherence function simplifies further to Eq.~(\ref{eq:Gamma_lambda}). It is this form of the decoherence function that we use in the main text.  Interestingly, displacing a reservoir state has no effect on the dephasing process. For example, this means that it does not matter for the dephasing process if either a mode is in a highly excited coherent state or the vacuum state. 

\section{Typical values of the dephasing susceptibility}\label{sec:typical_dephasing_susceptibility}
Here, we ask the question if there many $\bi-\bj$ such that $\gamma_{\bi-\bj}(\bk)\approx L^2$, given fixed values for $L$ and $\bk$. We show this is not the case:  as we go over all $(\bi,\bj)$, the values of $\gamma_{\bi-\bj}(\bk)$ are distributed according to a Gaussian that has a standard deviation that is at most $L/(2\pi)$. This means that, for a random $(\bi,\bj)$, $\gamma_{\bi-\bj}(\bk)$ is typically on the order of $L/(2\pi)$, or less.

To show this, fix $L$ and $\bk$, and consider the function $D_{\bi\bj} \defeq \sqrt{\gamma_{\bi-\bj}(\bk)} = || \sum_\ell (\bi_\ell - \bj_\ell) e^{i \bk \cdot \br_\ell}||$. Consider the frequency distribution of this function. This is a table that, per possible value $D_0$ of $D_{\bi\bj}$, shows the number of inputs $(\bi,\bj)$ such that $D_{\bi\bj}=D_0$. To obtain this distribution, we see $D$ as the distance from the origin of a random walker on the complex plane. The walker takes $L$ steps, where the $\ell$th step is given by $\bd_\ell e^{i \bk \cdot \br_\ell}$, with $\bd=\bi-\bj$. For the $\ell$th step, the walker has a probability 1/2 to make no step at all, a probability of 1/4 to take the step $+e^{i \bk \cdot \br_\ell}$, and a probability of 1/4 to take the step $-e^{i \bk \cdot \br_\ell}$. After $L$ steps, the walker is a distance $D_{\bi\bj}$ away from the origin of the complex plane. 

Naturally, the variance in the distances form the origin is largest if the walker is restricted to move on a single line, which happens if $\bk=\mathbf{0}$. Let us therefore put $\bk=\mathbf 0$, keeping in mind that, at worst, we are overestimating the variance of $D_{\bi\bj}$ for other values of $\bk$. For a 1D random walker that can take the steps $+1$ and $-1$ with equal probability, it is well-known that, after $L$ steps, the distribution of distances from the origin is well approximated by a Gaussian with standard deviation $\sqrt{2 L/\pi}$. In our situation, half of the time the 1D walker does not take a step at all. Therefore, the distribution of $D$ will be approximated by a Gaussian with variance $\sqrt{L/(2\pi)}$. Since $\gamma_{\bi-\bj}=(D_{\bi\bj})^2$, the distribution of $ \gamma_{\bi-\bj}$ over $(\bi,\bj)$ is approximated by a Gaussian with standard deviation $L/(2\pi)$. This means that for fixed $L$ and $\bk$, and given a random $(\bi,\bj)$, the decoherence function is, at most, typically on the order of $L/(2\pi)$. Additionally, it means that, if we are given a random $(\bi,\bj)$, where also the dimension $L$ of $\bi$ and $\bj$ is random but equal, the probability that $\gamma_{\bi-\bj}\geq\kappa L^2$ goes to zero as $L$ goes to infinity, for all $\bk$ and $\kappa>0$. 
\section{Explicit expressions for the vacuum contribution}\label{sec:explicit_expressions_for_the_vacuum_contribution}
Here, we derive the explicit solution of the vacuum part of the decoherence function, $\Gamma^{(vac)}_\bd(t)$, in the array model of Sec. \ref{sec:dephasing_susceptibility}. The assumptions are that the qubits form a linear array with spacing $a$, coupling to a one-dimensional reservoir via the single-reservoir dephasing Hamiltonian. The spectral density of the reservoir is assumed to be given by Eq.~(\ref{eq:spectral_density}). 

In principle, in the array model, $d=1$, but we will analytically extend our solutions to arbitrary $d$. After absorbing the integral over the solid angle, which in $d=1$ dimensions gives a factor of 2, into $\alpha_d$,  the vacuum decoherence function reads
\begin{align}\label{eq:Gamma_explicit}
\Gamma^{(vac)}_\bd(t)= \int_0^{\infty} \dd \om \, J(\om) \gamma_\bd(\om) \tau(t,\om),
\end{align}
with
\begin{align*}
J(\omega)&=\alpha_d \omega^d e^{-\om/\om_c},\\
\gamma_\bd(\om)&=\sum_{\ell m} \bd_\ell \bd_m \cos[\om a (\ell-m)],\\
\tau(t,\om)&=\frac{1-\cos(\omega t)}{\om^2}.
\end{align*}

In this section, we derive the full solution of Eq.~(\ref{eq:Gamma_explicit}). Additionally,  we derive simplified approximate solutions for the limits of infinitesimal and infinite time. For infinitesimal times, we find
\begin{align*}
\Gamma^{(vac)}_\bd(t)\approx\half \alpha_d \lVert \bd \rVert^2 \tilde \Gamma(1+d) \om_c^{d-1}(t\om_c)^2,
\end{align*}
where $\tilde \Gamma$ is the regular gamma function $\tilde \Gamma(j+1)=j!$, not to be confused with the decoherence function. In the infinite time limit, $\Gamma^{(vac)}_\bd(t)$ reaches a plateau for all $d>1$. For $d\geq 2$, we show that the height this plateau equals
\begin{equation*}
\lim_{t\to\infty} \Gamma^{(vac)}_{\bd}(t)\approx\alpha_d \lVert\bd\rVert^2 \tilde\Gamma{(d-1)}\om_c^{d-1}.
\end{equation*}
This result extends that of Sec.~\ref{sec:infinite_time_limit} for the current, specific setting. Note that, because $||\bd||^2\leq L$, the decoherence function scales at most linearly with $L$ in the limits of infinitesimal and infinite time, in accordance with the results in the main text. 

\subsection{General solution}
We start by rewriting $\gamma_\bd(\om)$ as 
\begin{align}\label{eq:gamma_computational}
\gamma_\bd(\om)= \sum_{r=0}^{L-1} f_{\bd r} \cos(a\om r),
\end{align}
where
\begin{align}\label{eq:definition_of_f}
f_{\bd r}=(2-\delta_{0r})\sum_{m=1}^{L-r}\bd_m \bd_{m+r}.    
\end{align}
Written this way, $\gamma_{\bd}(\om)$ is the cosine transform of $f_{\bd r}$.
For later reference, we note that for the states $\GHZ$ and $\GHZp$,
\begin{align*}
f_{\bd r }=\left\{\begin{array}{cc}L & :r=0 \\ 2(L-r)\zeta^r & :r>0\end{array}\right. ,
\end{align*}
where $\zeta=1$ for $\GHZ$ and $\zeta=-1$ for $\GHZp$.  

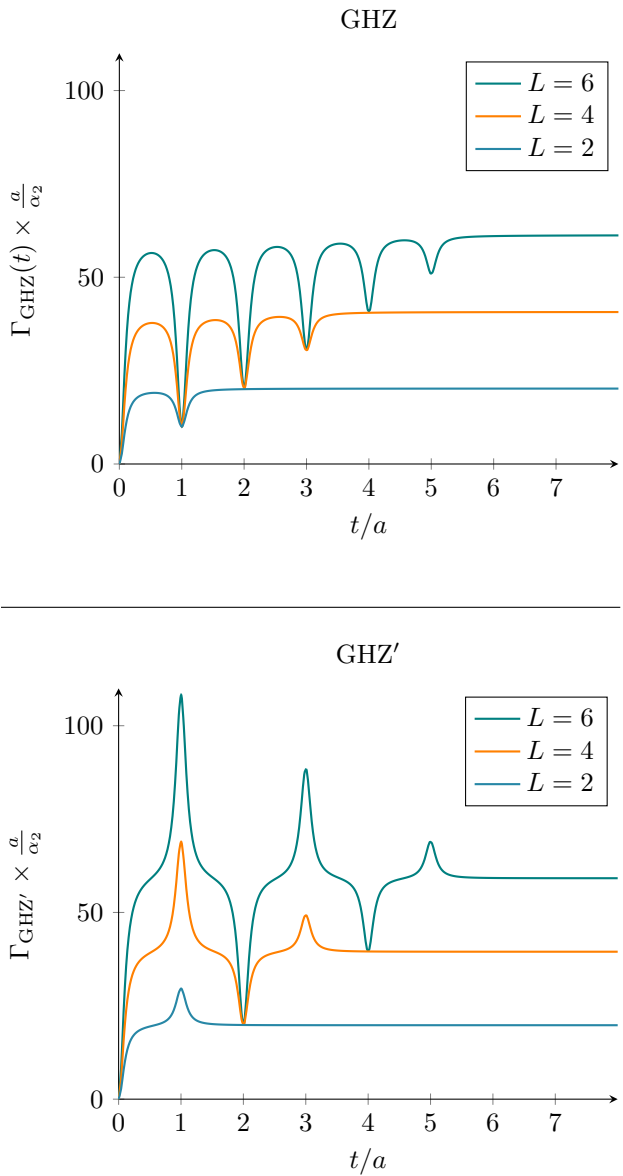
\begin{figure}
	\begin{tikzpicture}
	\begin{axis}[trig format plots=rad, cycle list name=exotic, width=\columnwidth, xtick={0,1,2,3,4,5,6,7}, ymin=0, title=GHZ, ymax=110,ytick={0,50,100},axis lines = left,xlabel=$t/a$, domain=-0:2*3.1415, ylabel=$\Gamma_\mrm{GHZ}(t)\times  \frac{a}{\al_2}$]
	\addplot+[thick, mark=none,smooth]
	coordinates {
		(0.0,0.0)(0.04020100502512563,8.342656854706036)(0.08040201005025126,23.53782005809017)(0.12060301507537688,35.50939146726281)(0.16080402010050251,43.184422153724846)(0.20100502512562815,47.96341670392259)(0.24120603015075376,51.003373417141894)(0.2814070351758794,52.995922057547965)(0.32160804020100503,54.33360813069991)(0.36180904522613067,55.241263253398905)(0.4020100502512563,55.85020198865234)(0.44221105527638194,56.23764205515089)(0.4824120603015075,56.44744940036065)(0.5226130653266332,56.50077637054177)(0.5628140703517588,56.400751253732004)(0.6030150753768845,56.13293729952483)(0.6432160804020101,55.66168146552882)(0.6834170854271356,54.92092332095618)(0.7236180904522613,53.79576573952321)(0.7638190954773869,52.087099668306216)(0.8040201005025126,49.444972332658324)(0.8442211055276382,45.252519634022185)(0.8844221105527639,38.49430206573824)(0.9246231155778895,28.059683280447842)(0.964824120603015,15.482438274279438)(1.0050251256281406,10.125433572508541)(1.0452261306532664,18.502488399565664)(1.085427135678392,31.109476160690875)(1.1256281407035176,40.61464913647063)(1.1658291457286432,46.65645889378927)(1.206030150753769,50.42789687734574)(1.2462311557788945,52.84037963114509)(1.2864321608040201,54.43146932649936)(1.3266331658291457,55.506177995294365)(1.3668341708542713,56.239739765832596)(1.407035175879397,56.734838061324936)(1.4472361809045227,57.05195409682347)(1.4874371859296482,57.225341770550806)(1.5276381909547738,57.2711868418681)(1.5678391959798996,57.19110644380758)(1.6080402010050252,56.972252612868026)(1.6482412060301508,56.584008259856965)(1.6884422110552764,55.96997117323208)(1.728643216080402,55.03197368761601)(1.7688442211055277,53.599432937263806)(1.8090452261306533,51.37184268964583)(1.849246231155779,47.8207320054383)(1.8894472361809045,42.094131904388284)(1.92964824120603,33.37864292919962)(1.9698492462311559,23.50107705998353)(2.0100502512562812,20.591601503452562)(2.050251256281407,28.27173028252512)(2.090452261306533,38.215010688568476)(2.130653266331658,45.438818170139704)(2.170854271356784,50.00039824309517)(2.21105527638191,52.855970653631104)(2.251256281407035,54.692537137764184)(2.291457286432161,55.911010490029724)(2.3316582914572863,56.73898029640027)(2.371859296482412,57.307649007244876)(2.412060301507538,57.694186467066984)(2.4522613065326633,57.94417045851096)(2.492462311557789,58.08341076604937)(2.5326633165829144,58.123974229469916)(2.5728643216080402,58.06671317618018)(2.613065326633166,57.90119292010208)(2.6532663316582914,57.60294159754842)(2.693467336683417,57.126923266782406)(2.7336683417085426,56.394561351032614)(2.7738693467336684,55.26885460642777)(2.814070351758794,53.5079065992891)(2.8542713567839195,50.687761818199654)(2.8944723618090453,46.142304934937904)(2.9346733668341707,39.34593548783693)(2.9748743718592965,32.18379680603801)(3.0150753768844223,31.0923995822193)(3.0552763819095476,37.4642777700248)(3.0954773869346734,44.760982437385685)(3.1356783919597992,49.900258913200155)(3.1758793969849246,53.131206884143396)(3.2160804020100504,55.16162442676381)(3.2562814070351758,56.475785295197)(3.2964824120603016,57.35399428601268)(3.3366834170854274,57.95560085735838)(3.3768844221105527,58.372858611550186)(3.4170854271356785,58.660311821571824)(3.457286432160804,58.85036983427351)(3.4974874371859297,58.96153712941202)(3.5376884422110555,59.00261985130758)(3.577889447236181,58.974491318521935)(3.6180904522613067,58.870024798852334)(3.658291457286432,58.672118884144815)(3.698492462311558,58.34900907738234)(3.7386934673366836,57.844922346701296)(3.778894472361809,57.06213617374091)(3.819095477386935,55.82765803818515)(3.85929648241206,53.83965681307875)(3.899497487437186,50.638680461530626)(3.9396984924623117,45.953385748499784)(3.979899497487437,41.423703050814694)(4.0201005025125625,41.44630136675811)(4.060301507537688,46.02150286559883)(4.100502512562814,50.753297158428325)(4.14070351758794,54.002456519454064)(4.180904522613066,56.04110450609898)(4.221105527638191,57.329591816540734)(4.261306532663316,58.17081850472416)(4.301507537688442,58.7390854550022)(4.341708542713568,59.13375828502336)(4.381909547738694,59.412722926643276)(4.42211055276382,59.61052296945284)(4.4623115577889445,59.74799782924543)(4.50251256281407,59.83741860027689)(4.542713567839196,59.88516498622357)(4.582914572864322,59.89292659563753)(4.623115577889448,59.85782536749216)(4.663316582914573,59.77145526641438)(4.703517587939698,59.61742140175569)(4.743718592964824,59.36633394473338)(4.78391959798995,58.96613737501351)(4.824120603015076,58.32424337777062)(4.864321608040201,57.27992055065494)(4.9045226130653266,55.59782032183338)(4.944723618090452,53.19541797781187)(4.984924623115578,51.10059100228798)(5.025125628140704,51.49325432358708)(5.065326633165829,53.90965419911983)(5.105527638190955,56.204722576156676)(5.1457286432160805,57.750545122363405)(5.185929648241206,58.72284283153456)(5.226130653266332,59.344256829776874)(5.266331658291457,59.756825841304355)(5.306532663316583,60.04206755434379)(5.346733668341709,60.246766029475054)(5.386934673366834,60.39855592265595)(5.42713567839196,60.514354390699374)(5.467336683417085,60.60489185145756)(5.507537688442211,60.6772024387825)(5.547738693467337,60.73603668600982)(5.5879396984924625,60.78468967634404)(5.628140703517588,60.82550215552392)(5.668341708542713,60.860172830825405)(5.708542713567839,60.88995833730777)(5.748743718592965,60.91580447572586)(5.788944723618091,60.938434300936386)(5.8291457286432165,60.9584084703312)(5.869346733668341,60.976167367225244)(5.909547738693467,60.99206100960273)(5.949748743718593,61.00637062136344)(5.989949748743719,61.019324415961435)(6.030150753768845,61.031109299656116)(6.07035175879397,61.04187965644319)(6.110552763819095,61.05176401786262)(6.150753768844221,61.0608701807833)(6.190954773869347,61.06928917319445)(6.231155778894473,61.077098355713304)(6.2713567839195985,61.08436386813429)(6.311557788944723,61.09114257496643)(6.351758793969849,61.09748362432879)(6.391959798994975,61.10342970598458)(6.432160804020101,61.109018073427514)(6.472361809045227,61.11428137956181)(6.5125628140703515,61.119248364085685)(6.552763819095477,61.123944422118115)(6.592964824120603,61.1283920771277)(6.633165829145729,61.13261137628788)(6.673366834170855,61.13662022259449)(6.71356783919598,61.14043465515535)(6.7537688442211055,61.14406908678589)(6.793969849246231,61.147536506263194)(6.834170854271357,61.15084865118838)(6.874371859296483,61.15401615629717)(6.914572864321608,61.157048681173464)(6.954773869346734,61.1599550206142)(6.994974874371859,61.16274320032367)(7.035175879396985,61.16542056015602)(7.075376884422111,61.16799382675106)(7.115577889447236,61.170469177103534)(7.155778894472362,61.17285229435746)(7.1959798994974875,61.17514841691098)(7.236180904522613,61.177362381748424)(7.276381909547739,61.17949866277599)(7.316582914572864,61.18156140482056)(7.35678391959799,61.18355445385337)(7.396984924623116,61.185481383919765)(7.437185929648241,61.18734552118657)(7.477386934673367,61.189149965461354)(7.517587939698492,61.1908976094892)(7.557788944723618,61.192591156290874)(7.597989949748744,61.19423313477107)(7.63819095477387,61.19582591379552)(7.678391959798995,61.197371714910275)(7.71859296482412,61.198872623853696)(7.758793969849246,61.20033060099385)(7.798994974874372,61.201747490806966)(7.839195979899498,61.20312503049829)(7.8793969849246235,61.204464857855925)(7.919597989949749,61.2057685184155)(7.959798994974874,61.207037472006604)(8.0,61.20827309874206)
	};		
	\addlegendentry{$L=6$}
	\addplot+[thick, mark=none,smooth]
	coordinates {
		(0.0,0.0)(0.04020100502512563,5.562142821680541)(0.08040201005025126,15.6933766028385)(0.12060301507537688,23.67633390316061)(0.16080402010050251,28.79576988487828)(0.20100502512562815,31.985435480601648)(0.24120603015075376,34.016776526542436)(0.2814070351758794,35.35103400205163)(0.32160804020100503,36.25011246989399)(0.36180904522613067,36.86417818647147)(0.4020100502512563,37.281149924920065)(0.44221105527638194,37.553018306893236)(0.4824120603015075,37.70973544717136)(0.5226130653266332,37.76639974196467)(0.5628140703517588,37.72653006724411)(0.6030150753768845,37.58261440696233)(0.6432160804020101,37.3140863367556)(0.6834170854271356,36.881915746809035)(0.7236180904522613,36.21761948095624)(0.7638190954773869,35.2020815054052)(0.8040201005025126,33.6256046719902)(0.8442211055276382,31.118291680665457)(0.8844221105527639,27.071062061227124)(0.9246231155778895,20.817690730398645)(0.964824120603015,13.278581682659382)(1.0050251256281406,10.071583199839255)(1.0452261306532664,15.105110668090784)(1.085427135678392,22.67681315466076)(1.1256281407035176,28.387773980450877)(1.1658291457286432,32.02120929442786)(1.206030150753769,34.29307991224131)(1.2462311557788945,35.75043184025672)(1.2864321608040201,36.71604093399926)(1.3266331658291457,37.373212254975314)(1.3668341708542713,37.82746701827499)(1.407035175879397,38.14091292674903)(1.4472361809045227,38.350503512083264)(1.4874371859296482,38.47769914791804)(1.5276381909547738,38.53348077592117)(1.5678391959798996,38.52063845552265)(1.6080402010050252,38.43414421662321)(1.6482412060301508,38.25969741418322)(1.6884422110552764,37.969841605472716)(1.728643216080402,37.51605554953772)(1.7688442211055277,36.81348244862576)(1.8090452261306533,35.71221284874257)(1.849246231155779,33.948289891278)(1.8894472361809045,31.095961319064873)(1.92964824120603,26.748727147631637)(1.9698492462311559,21.820171373279106)(2.0100502512562812,20.375542625627755)(2.050251256281407,24.225757574098)(2.090452261306533,29.207751301535684)(2.130653266331658,32.830382049064255)(2.170854271356784,35.12246026304331)(2.21105527638191,36.56231031023952)(2.251256281407035,37.493686509020854)(2.291457286432161,38.11735369745831)(2.3316582914572863,38.547492020576364)(2.371859296482412,38.850195829228184)(2.412060301507538,39.06469554358733)(2.4522613065326633,39.2146439929708)(2.492462311557789,39.31412879568282)(2.5326633165829144,39.37084133276072)(2.5728643216080402,39.38759176345437)(2.613065326633166,39.36268704837697)(2.6532663316582914,39.28926237104998)(2.693467336683417,39.15326304652575)(2.7336683417085426,38.92922172866681)(2.7738693467336684,38.57203243061916)(2.814070351758794,38.00150743975661)(2.8542713567839195,37.07668912593768)(2.8944723618090453,35.57584028792517)(2.9346733668341707,33.32402024711304)(2.9748743718592965,30.949841718828143)(3.0150753768844223,30.59902339498354)(3.0552763819095476,32.735951888875185)(3.0954773869346734,35.181355810142605)(3.1356783919597992,36.90804123470325)(3.1758793969849246,37.99928931667367)(3.2160804020100504,38.69131739764063)(3.2562814070351758,39.14588807032126)(3.2964824120603016,39.4568497055424)(3.3366834170854274,39.67783797736916)(3.3768844221105527,39.8402718876436)(3.4170854271356785,39.96321145500987)(3.457286432160804,40.05864163926122)(3.4974874371859297,40.134356648674064)(3.5376884422110555,40.19558271632626)(3.577889447236181,40.245921663230774)(3.6180904522613067,40.2879174466483)(3.658291457286432,40.32340672448792)(3.698492462311558,40.35374186049274)(3.7386934673366836,40.37993641360683)(3.778894472361809,40.40276226405717)(3.819095477386935,40.422815825340685)(3.85929648241206,40.440564052217475)(3.899497487437186,40.45637697286159)(3.9396984924623117,40.47055106329574)(3.979899497487437,40.4833262907318)(4.0201005025125625,40.49489871012942)(4.060301507537688,40.50542989144801)(4.100502512562814,40.51505405727791)(4.14070351758794,40.523883545445855)(4.180904522613066,40.532013031803956)(4.221105527638191,40.53952282527495)(4.261306532663316,40.54648146156674)(4.301507537688442,40.5529477616327)(4.341708542713568,40.55897247795503)(4.381909547738694,40.56459962074725)(4.42211055276382,40.569867533617405)(4.4623115577889445,40.57480977165468)(4.50251256281407,40.579455822603144)(4.542713567839196,40.58383170258166)(4.582914572864322,40.58796045086643)(4.623115577889448,40.59186254297195)(4.663316582914573,40.595556237222645)(4.703517587939698,40.59905786688818)(4.743718592964824,40.6023820875329)(4.78391959798995,40.60554208733758)(4.824120603015076,40.6085497666631)(4.864321608040201,40.61141589194982)(4.9045226130653266,40.61415022810997)(4.944723618090452,40.616761652823406)(4.984924623115578,40.61925825554596)(5.025125628140704,40.62164742355461)(5.065326633165829,40.62393591696083)(5.105527638190955,40.62612993430277)(5.1457286432160805,40.628235170064706)(5.185929648241206,40.630256865257515)(5.226130653266332,40.63219985201546)(5.266331658291457,40.634068593018355)(5.306532663316583,40.6358672164255)(5.346733668341709,40.637599546906195)(5.386934673366834,40.639269133266225)(5.42713567839196,40.6408792730981)(5.467336683417085,40.642433034823114)(5.507537688442211,40.64393327744124)(5.547738693467337,40.64538266826323)(5.5879396984924625,40.64678369886119)(5.628140703517588,40.64813869944371)(5.668341708542713,40.649449851834376)(5.708542713567839,40.65071920120994)(5.748743718592965,40.651948666734455)(5.788944723618091,40.65314005120915)(5.8291457286432165,40.65429504984312)(5.869346733668341,40.655415258236765)(5.909547738693467,40.65650217966017)(5.949748743718593,40.657557231697695)(5.989949748743719,40.65858175232279)(6.030150753768845,40.65957700545943)(6.07035175879397,40.66054418608039)(6.110552763819095,40.66148442488661)(6.150753768844221,40.662398792607924)(6.190954773869347,40.66328830396031)(6.231155778894473,40.66415392129138)(6.2713567839195985,40.66499655794269)(6.311557788944723,40.66581708135435)(6.351758793969849,40.66661631593454)(6.391959798994975,40.66739504571497)(6.432160804020101,40.66815401681039)(6.472361809045227,40.668893939699274)(6.5125628140703515,40.669615491340366)(6.552763819095477,40.67031931713919)(6.592964824120603,40.67100603277653)(6.633165829145729,40.67167622591032)(6.673366834170855,40.67233045776104)(6.71356783919598,40.672969264589874)(6.7537688442211055,40.67359315907802)(6.793969849246231,40.674202631615)(6.834170854271357,40.67479815150261)(6.874371859296483,40.67538016808132)(6.914572864321608,40.675949111784554)(6.954773869346734,40.67650539512656)(6.994974874371859,40.677049413628374)(7.035175879396985,40.67758154668675)(7.075376884422111,40.6781021583899)(7.115577889447236,40.67861159828374)(7.155778894472362,40.67911020209253)(7.1959798994974875,40.67959829239668)(7.236180904522613,40.68007617927078)(7.276381909547739,40.68054416088465)(7.316582914572864,40.681002524069946)(7.35678391959799,40.68145154485429)(7.396984924623116,40.68189148896562)(7.437185929648241,40.68232261230816)(7.477386934673367,40.68274516141219)(7.517587939698492,40.68315937385924)(7.557788944723618,40.68356547868418)(7.597989949748744,40.68396369675573)(7.63819095477387,40.68435424113675)(7.678391959798995,40.684737317425544)(7.71859296482412,40.68511312407926)(7.758793969849246,40.685481852720656)(7.798994974874372,40.68584368842911)(7.839195979899498,40.6861988100167)(7.8793969849246235,40.686547390290585)(7.919597989949749,40.68688959630208)(7.959798994974874,40.68722558958349)(8.0,40.68755552637333)
	};
	\addlegendentry{$L=4$}
	\addplot+[thick, mark=none,smooth]
	coordinates {
		(0.0,0.0)(0.04020100502512563,2.7816077474184286)(0.08040201005025126,7.848848922677751)(0.12060301507537688,11.843086607762272)(0.16080402010050251,14.406779753861487)(0.20100502512562815,16.00692521450514)(0.24120603015075376,17.029415811312774)(0.2814070351758794,17.705103057369225)(0.32160804020100503,18.16524975669734)(0.36180904522613067,18.485355849336507)(0.4020100502512563,18.709943216026172)(0.44221105527638194,18.865774124640172)(0.4824120603015075,18.96888543745131)(0.5226130653266332,19.028320007533218)(0.5628140703517588,19.04798552079444)(0.6030150753768845,19.027292719380156)(0.6432160804020101,18.960759607550877)(0.6834170854271356,18.83638397673884)(0.7236180904522613,18.632093971488857)(0.7638190954773869,18.308763636129843)(0.8040201005025126,17.7969482115272)(0.8442211055276382,16.97371363411148)(0.8844221105527639,15.636334551585783)(0.9246231155778895,13.562992833430787)(0.964824120603015,11.060716727129241)(1.0050251256281406,10.002331044698364)(1.0452261306532664,11.690841572971843)(1.085427135678392,14.225666684101757)(1.1256281407035176,16.140713715742265)(1.1658291457286432,17.36395562797425)(1.206030150753769,18.134313995702442)(1.2462311557788945,18.634454727323636)(1.2864321608040201,18.97235672845592)(1.3266331658291457,19.20960623667144)(1.3668341708542713,19.38199831078747)(1.407035175879397,19.511050101404734)(1.4472361809045227,19.61017080161262)(1.4874371859296482,19.688008504865017)(1.5276381909547738,19.750318150486308)(1.5678391959798996,19.801038718907733)(1.6080402010050252,19.842934938245456)(1.6482412060301508,19.877991458149072)(1.6884422110552764,19.90766191799587)(1.728643216080402,19.933030446355684)(1.7688442211055277,19.95491887335386)(1.8090452261306533,19.973959445606404)(1.849246231155779,19.99064511643467)(1.8894472361809045,20.005364952881738)(1.92964824120603,20.018429474582085)(1.9698492462311559,20.030089061305258)(2.0100502512562812,20.04054751102059)(2.050251256281407,20.049972154093957)(2.090452261306533,20.05850148785442)(2.130653266331658,20.066251002814123)(2.170854271356784,20.073317674322244)(2.21105527638191,20.079783458334827)(2.251256281407035,20.08571803630665)(2.291457286432161,20.091180988430583)(2.3316582914572863,20.096223527705206)(2.371859296482412,20.100889893722382)(2.412060301507538,20.10521848067625)(2.4522613065326633,20.109242756209895)(2.492462311557789,20.112992014477243)(2.5326633165829144,20.116491996913307)(2.5728643216080402,20.119765406763236)(2.613065326633166,20.122832337773385)(2.6532663316582914,20.12571063312956)(2.693467336683417,20.128416187403257)(2.7336683417085426,20.130963201689916)(2.7738693467336684,20.13336440011242)(2.814070351758794,20.135631214285397)(2.8542713567839195,20.13777394108996)(2.8944723618090453,20.139801878119247)(2.9346733668341707,20.14172344036599)(2.9748743718592965,20.14354626108975)(3.0150753768844223,20.145277279291157)(3.0552763819095476,20.146922815806853)(3.0954773869346734,20.14848863970237)(3.1356783919597992,20.14998002636574)(3.1758793969849246,20.151401808478354)(3.2160804020100504,20.15275842085475)(3.2562814070351758,20.154053939988525)(3.2964824120603016,20.155292119014785)(3.3366834170854274,20.156476418692897)(3.3768844221105527,20.157610034924787)(3.4170854271356785,20.158695923249567)(3.457286432160804,20.159736820692675)(3.4974874371859297,20.16073526529486)(3.5376884422110555,20.161693613601713)(3.577889447236181,20.162614056356414)(3.6180904522613067,20.163498632606153)(3.658291457286432,20.16434924240504)(3.698492462311558,20.16516765827287)(3.7386934673366836,20.165955535548648)(3.778894472361809,20.166714421760705)(3.819095477386935,20.167445765119858)(3.85929648241206,20.16815092222949)(3.899497487437186,20.16883116509494)(3.9396984924623117,20.169487687504937)(3.979899497487437,20.17012161084938)(4.0201005025125625,20.170733989430406)(4.060301507537688,20.171325815317065)(4.100502512562814,20.171898022788614)(4.14070351758794,20.17245149240615)(4.180904522613066,20.172987054748173)(4.221105527638191,20.173505493841848)(4.261306532663316,20.17400755031828)(4.301507537688442,20.1744939243172)(4.341708542713568,20.17496527816384)(4.381909547738694,20.175422238838465)(4.42211055276382,20.17586540025693)(4.4623115577889445,20.17629532537875)(4.50251256281407,20.176712548157738)(4.542713567839196,20.177117575348575)(4.582914572864322,20.177510888181505)(4.623115577889448,20.177892943916245)(4.663316582914573,20.178264177284973)(4.703517587939698,20.178625001833556)(4.743718592964824,20.178975811169217)(4.78391959798995,20.17931698012205)(4.824120603015076,20.17964886582729)(4.864321608040201,20.179971808734464)(4.9045226130653266,20.180286133549103)(4.944723618090452,20.180592150112165)(4.984924623115578,20.180890154221952)(5.025125628140704,20.181180428402694)(5.065326633165829,20.1814632426239)(5.105527638190955,20.181738854974025)(5.1457286432160805,20.18200751229179)(5.185929648241206,20.182269450758163)(5.226130653266332,20.182524896451895)(5.266331658291457,20.182774065871097)(5.306532663316583,20.183017166423298)(5.346733668341709,20.1832543968861)(5.386934673366834,20.183485947840552)(5.42713567839196,20.183712002078938)(5.467336683417085,20.183932734988925)(5.507537688442211,20.184148314915387)(5.547738693467337,20.184358903501632)(5.5879396984924625,20.18456465601122)(5.628140703517588,20.184765721631653)(5.668341708542713,20.184962243761202)(5.708542713567839,20.185154360279835)(5.748743718592965,20.185342203805295)(5.788944723618091,20.185525901935286)(5.8291457286432165,20.185705577476625)(5.869346733668341,20.18588134866207)(5.909547738693467,20.186053329355747)(5.949748743718593,20.186221629247758)(5.989949748743719,20.186386354038554)(6.030150753768845,20.186547605613914)(6.07035175879397,20.186705482210833)(6.110552763819095,20.186860078575023)(6.150753768844221,20.187011486110404)(6.190954773869347,20.187159793021184)(6.231155778894473,20.18730508444679)(6.2713567839195985,20.187447442590155)(6.311557788944723,20.187586946839804)(6.351758793969849,20.18772367388584)(6.391959798994975,20.187857697830538)(6.432160804020101,20.187989090293417)(6.472361809045227,20.188117920511537)(6.5125628140703515,20.188244255434856)(6.552763819095477,20.188368159817294)(6.592964824120603,20.188489696303424)(6.633165829145729,20.18860892551124)(6.673366834170855,20.188725906111085)(6.71356783919598,20.188840694901018)(6.7537688442211055,20.188953346878705)(6.793969849246231,20.189063915310186)(6.834170854271357,20.189172451795436)(6.874371859296483,20.189279006331176)(6.914572864321608,20.18938362737077)(6.954773869346734,20.189486361881592)(6.994974874371859,20.189587255399875)(7.035175879396985,20.189686352083204)(7.075376884422111,20.189783694760735)(7.115577889447236,20.1898793249813)(7.155778894472362,20.189973283059434)(7.1959798994974875,20.19006560811954)(7.236180904522613,20.19015633813813)(7.276381909547739,20.19024550998435)(7.316582914572864,20.190333159458845)(7.35678391959799,20.190419321330996)(7.396984924623116,20.190504029374654)(7.437185929648241,20.190587316402457)(7.477386934673367,20.1906692142987)(7.517587939698492,20.190749754050945)(7.557788944723618,20.19082896578037)(7.597989949748744,20.19090687877087)(7.63819095477387,20.19098352149707)(7.678391959798995,20.191058921651237)(7.71859296482412,20.19113310616908)(7.758793969849246,20.19120610125465)(7.798994974874372,20.191277932404244)(7.839195979899498,20.191348624429335)(7.8793969849246235,20.19141820147877)(7.919597989949749,20.19148668705996)(7.959798994974874,20.19155410405943)(8.0,20.191620474762527)
	};
	\addlegendentry{$L=2$}
	\end{axis}	
	
	\end{tikzpicture}
	\vspace{1em}
	\hrule
	\vspace{1em}
	\begin{tikzpicture}
	\begin{axis}[trig format plots=rad, cycle list name=exotic, ymin=0, ymax=110, width=\columnwidth, xtick={0,1,2,3,4,5,6,7}, title=GHZ$'$, ytick={0,50,100},axis lines = left,xlabel=$t/a$, ylabel=$\Gamma_\mrm{GHZ'}\times  \frac{a}{\al_2}$, domain=-0:2*3.1415 ]
	\addplot+[thick, mark=none,smooth]
	coordinates {
		(0.0,0.0)(0.04020100502512563,8.352136117929508)(0.08040201005025126,23.576032791555946)(0.12060301507537688,35.59649830205252)(0.16080402010050251,43.34216120736955)(0.20100502512562815,48.215882784837504)(0.24120603015075376,51.377965998500635)(0.2814070351758794,53.5245437030599)(0.32160804020100503,55.05422843107326)(0.36180904522613067,56.20001596618793)(0.4020100502512563,57.10426684243976)(0.44221105527638194,57.859298258906996)(0.4824120603015075,58.52991210630554)(0.5226130653266332,59.166807428709376)(0.5628140703517588,59.815697851169666)(0.6030150753768845,60.52504032476596)(0.6432160804020101,61.35469286391616)(0.6834170854271356,62.38825416737599)(0.7236180904522613,63.75352385797559)(0.7638190954773869,65.65921756584856)(0.8040201005025126,68.46250000939173)(0.8442211055276382,72.78566858931981)(0.8844221105527639,79.64824327163753)(0.9246231155778895,90.16388370997512)(0.964824120603015,102.80098506668176)(1.0050251256281406,108.1981347611397)(1.0452261306532664,99.84233661792352)(1.085427135678392,87.23795211413072)(1.1256281407035176,77.71637991998351)(1.1658291457286432,71.63820101758103)(1.206030150753769,67.80876030473354)(1.2462311557788945,65.31415030395387)(1.2864321608040201,63.61328917759977)(1.3266331658291457,62.39631693339885)(1.3668341708542713,61.48138133502718)(1.407035175879397,60.75675576457789)(1.4472361809045227,60.14950757453497)(1.4874371859296482,59.608020681772054)(1.5276381909547738,59.091489225146944)(1.5678391959798996,58.5626573083261)(1.6080402010050252,57.98151639084116)(1.6482412060301508,57.298069622531436)(1.6884422110552764,56.44183483424651)(1.728643216080402,55.30425882646694)(1.7688442211055277,53.70700544773461)(1.8090452261306533,51.34373098576209)(1.849246231155779,47.68155188128476)(1.8894472361809045,41.865171124192564)(1.92964824120603,33.078708823565414)(1.9698492462311559,23.14718000953676)(2.0100502512562812,20.199527666934703)(2.050251256281407,27.856544624872342)(2.090452261306533,37.791524141929294)(2.130653266331658,45.02204943822931)(2.170854271356784,49.606060844844286)(2.21105527638191,52.501019384503024)(2.251256281407035,54.39581590962211)(2.291457286432161,55.694056998681475)(2.3316582914572863,56.627064876115924)(2.371859296482412,57.33116006677855)(2.412060301507538,57.8905311957333)(2.4522613065326633,58.360451316539006)(2.492462311557789,58.78029122902366)(2.5326633165829144,59.18140844191243)(2.5728643216080402,59.5926726591457)(2.613065326633166,60.04535244617252)(2.6532663316582914,60.57883093895235)(2.693467336683417,61.249010544493224)(2.7336683417085426,62.142508452225194)(2.7738693467336684,63.402349628678245)(2.814070351758794,65.27498463764077)(2.8542713567839195,68.18782976316372)(2.8944723618090453,72.80962947828029)(2.9346733668341707,79.66796493350232)(2.9748743718592965,86.87915278510874)(3.0150753768844223,88.00770889666966)(3.0552763819095476,81.66174873442546)(3.0954773869346734,74.38002450676252)(3.1356783919597992,69.2447564845487)(3.1758793969849246,66.00645911070232)(3.2160804020100504,63.95655864154931)(3.2562814070351758,62.6095477159333)(3.2964824120603016,61.6833226848528)(3.3366834170854274,61.01600819949161)(3.3768844221105527,60.51185680464789)(3.4170854271356785,60.1115025945206)(3.457286432160804,59.7758430022124)(3.4974874371859297,59.47697213028964)(3.5376884422110555,59.19265832732355)(3.577889447236181,58.90245493203745)(3.6180904522613067,58.584239546493855)(3.658291457286432,58.210141209941206)(3.698492462311558,57.74051873231142)(3.7386934673366836,57.113746860424044)(3.778894472361809,56.22769706296498)(3.819095477386935,54.90602436105175)(3.85929648241206,52.84432761703418)(3.899497487437186,49.58117753534995)(3.9396984924623117,44.84371246502879)(3.979899497487437,40.2707084902308)(4.0201005025125625,40.257978081562214)(4.060301507537688,44.805239626956464)(4.100502512562814,49.51610363032361)(4.14070351758794,52.75117435925454)(4.180904522613066,54.78261854511316)(4.221105527638191,56.07105566010811)(4.261306532663316,56.919911301607236)(4.301507537688442,57.5043212392886)(4.341708542713568,57.924877184525016)(4.381909547738694,58.241206424432356)(4.42211055276382,58.49029639237532)(4.4623115577889445,58.696417796578864)(4.50251256281407,58.87670119880595)(4.542713567839196,59.04451029978493)(4.582914572864322,59.211771872842775)(4.623115577889448,59.39098099789582)(4.663316582914573,59.59748196415177)(4.703517587939698,59.852771460038724)(4.743718592964824,60.190055410911434)(4.78391959798995,60.664288078655055)(4.824120603015076,61.37029063113547)(4.864321608040201,62.47053635447934)(4.9045226130653266,64.20174850709284)(4.944723618090452,66.64754825452941)(4.984924623115578,68.78094119945565)(5.025125628140704,68.42273006458257)(5.065326633165829,66.03725665826913)(5.105527638190955,63.7700746330162)(5.1457286432160805,62.24950218594766)(5.185929648241206,61.300156455997374)(5.226130653266332,60.69968118371707)(5.266331658291457,60.30627903911412)(5.306532663316583,60.03863801134008)(5.346733668341709,59.850150188016)(5.386934673366834,59.71333250367304)(5.42713567839196,59.61139885505168)(5.467336683417085,59.5337326131236)(5.507537688442211,59.47339884684864)(5.547738693467337,59.42573384480806)(5.5879396984924625,59.38751878967194)(5.628140703517588,59.35648016327866)(5.668341708542713,59.330978715973345)(5.708542713567839,59.3098105650916)(5.748743718592965,59.29207685471551)(5.788944723618091,59.27709642720546)(5.8291457286432165,59.26434612063891)(5.869346733668341,59.25341919641231)(5.909547738693467,59.24399590223411)(5.949748743718593,59.23582230615576)(5.989949748743719,59.228694862261946)(6.030150753768845,59.222449009478346)(6.07035175879397,59.21695064861928)(6.110552763819095,59.21208970046104)(6.150753768844221,59.207775186736086)(6.190954773869347,59.20393143820432)(6.231155778894473,59.2004951456105)(6.2713567839195985,59.19741304717249)(6.311557788944723,59.19464010116642)(6.351758793969849,59.192138031371144)(6.391959798994975,59.18987416140901)(6.432160804020101,59.18782047461821)(6.472361809045227,59.18595285124507)(6.5125628140703515,59.184250445984084)(6.552763819095477,59.18269517730924)(6.592964824120603,59.18127130638236)(6.633165829145729,59.179965088147554)(6.673366834170855,59.1787644809103)(6.71356783919598,59.17765890354287)(6.7537688442211055,59.17663903166272)(6.793969849246231,59.17569662585062)(6.834170854271357,59.174824386324985)(6.874371859296483,59.17401582955447)(6.914572864321608,59.173265183135186)(6.954773869346734,59.17256729593321)(6.994974874371859,59.17191756103252)(7.035175879396985,59.17131184946284)(7.075376884422111,59.170746453032976)(7.115577889447236,59.17021803488074)(7.155778894472362,59.1697235865823)(7.1959798994974875,59.169260390854824)(7.236180904522613,59.168825989041764)(7.276381909547739,59.16841815269924)(7.316582914572864,59.16803485870897)(7.35678391959799,59.16767426743079)(7.396984924623116,59.167334703482254)(7.437185929648241,59.167014638794114)(7.477386934673367,59.16671267764163)(7.517587939698492,59.166427543395784)(7.557788944723618,59.16615806677454)(7.597989949748744,59.16590317540482)(7.63819095477387,59.165661884532824)(7.678391959798995,59.16543328874208)(7.71859296482412,59.165216554557226)(7.758793969849246,59.16501091382828)(7.798994974874372,59.164815657803985)(7.839195979899498,59.16463013181335)(7.8793969849246235,59.16445373048725)(7.919597989949749,59.16428589345758)(7.959798994974874,59.1641261014819)(8.0,59.16397387294604)
	};		
	\addlegendentry{$L=6$}
	\addplot+[thick, mark=none,smooth]
	coordinates {
		(0.0,0.0)(0.04020100502512563,5.567809431185489)(0.08040201005025126,15.716220378458102)(0.12060301507537688,23.72840904319223)(0.16080402010050251,28.890076723165333)(0.20100502512562815,32.13638786298695)(0.24120603015075376,34.24077044525091)(0.2814070351758794,35.667166496331)(0.32160804020100503,36.68111982697039)(0.36180904522613067,37.437694057145286)(0.4020100502512563,38.03143422898035)(0.44221105527638194,38.52338915134358)(0.4824120603015075,38.95607080460908)(0.5226130653266332,39.36230369020982)(0.5628140703517588,39.771155743958076)(0.6030150753768845,40.21284878138554)(0.6432160804020101,40.72412035538607)(0.6834170854271356,41.35573288611561)(0.7236180904522613,42.18481800156734)(0.7638190954773869,43.33695086464656)(0.8040201005025126,45.02670082761553)(0.8442211055276382,47.6276630731451)(0.8844221105527639,51.751727931691676)(0.9246231155778895,58.06724252061687)(0.964824120603015,65.65537695897947)(1.0050251256281406,68.89940544813211)(1.0452261306532664,63.891646799216424)(1.085427135678392,56.334834781566734)(1.1256281407035176,50.62793150178593)(1.1658291457286432,46.98742068603292)(1.206030150753769,44.69666399074812)(1.2462311557788945,43.207500462044486)(1.2864321608040201,42.19550550334133)(1.3266331658291457,41.475045201512856)(1.3668341708542713,40.9373712282876)(1.407035175879397,40.515925277284886)(1.4472361809045227,40.1675899689002)(1.4874371859296482,39.86227807072161)(1.5276381909547738,39.57674641777893)(1.5678391959798996,39.290428009773315)(1.6080402010050252,38.98196047038252)(1.6482412060301508,38.62537156199295)(1.6884422110552764,38.18470641331155)(1.728643216080402,37.60515401836704)(1.7688442211055277,36.797137609892076)(1.8090452261306533,35.60716576273019)(1.849246231155779,33.76854067490226)(1.8894472361809045,30.853403002006754)(1.92964824120603,26.453633534184963)(1.9698492462311559,21.481581606283488)(2.0100502512562812,20.001573101866033)(2.050251256281407,23.82386402221722)(2.090452261306533,28.78495886340954)(2.130653266331658,32.39349831514241)(2.170854271356784,34.67828498337829)(2.21105527638191,36.11785579893266)(2.251256281407035,37.056424704901424)(2.291457286432161,37.69550986554076)(2.3316582914572863,38.15041203197344)(2.371859296482412,38.488825374457406)(2.412060301507538,38.75222820237743)(2.4522613065326633,38.96742554309251)(2.492462311557789,39.15295601440607)(2.5326633165829144,39.32288051829032)(2.5728643216080402,39.489294965278035)(2.613065326633166,39.6643602648214)(2.6532663316582914,39.86245479433607)(2.693467336683417,40.103132742931635)(2.7336683417085426,40.41595802793828)(2.7738693467336684,40.849136774371516)(2.814070351758794,41.48523654056143)(2.8542713567839195,42.467065725360534)(2.8944723618090453,44.01781796650115)(2.9346733668341707,46.313592529877326)(2.9748743718592965,48.726707545913555)(3.0150753768844223,49.11219906204516)(3.0552763819095476,47.00629760450456)(3.0954773869346734,44.588783152144146)(3.1356783919597992,42.887271815803146)(3.1758793969849246,41.81883528344217)(3.2160804020100504,41.14755296848918)(3.2562814070351758,40.71191334006768)(3.2964824120603016,40.418281610889174)(3.3366834170854274,40.21320475418788)(3.3768844221105527,40.06542072629693)(3.4170854271356785,39.95600501440725)(3.457286432160804,39.87309015489114)(3.4974874371859297,39.80898424010531)(3.5376884422110555,39.75855045544412)(3.577889447236181,39.718265428111096)(3.6180904522613067,39.68565426694682)(3.658291457286432,39.65894132359811)(3.698492462311558,39.63682829882091)(3.7386934673366836,39.61834968916803)(3.778894472361809,39.602776452501715)(3.819095477386935,39.58955046809634)(3.85929648241206,39.57823910143899)(3.899497487437186,39.56850316201938)(3.9396984924623117,39.560073949424456)(3.979899497487437,39.552736572151495)(4.0201005025125625,39.54631766392785)(4.060301507537688,39.54067622759465)(4.100502512562814,39.53569673313042)(4.14070351758794,39.53128386044234)(4.180904522613066,39.52735845608878)(4.221105527638191,39.52385439553429)(4.261306532663316,39.520716127620574)(4.301507537688442,39.51789673779559)(4.341708542713568,39.515356409244596)(4.381909547738694,39.51306119171715)(4.42211055276382,39.51098201012049)(4.4623115577889445,39.50909386129609)(4.50251256281407,39.507375159499695)(4.542713567839196,39.50580720014317)(4.582914572864322,39.50437371815984)(4.623115577889448,39.503060522516435)(4.663316582914573,39.50185519233619)(4.703517587939698,39.500746823131884)(4.743718592964824,39.49972581399463)(4.78391959798995,39.498783688413994)(4.824120603015076,39.497912942838255)(4.864321608040201,39.49710691821353)(4.9045226130653266,39.496359690634755)(4.944723618090452,39.49566597795482)(4.984924623115578,39.495021059767865)(5.025125628140704,39.494420708641336)(5.065326633165829,39.49386113084168)(5.105527638190955,39.493338915098704)(5.1457286432160805,39.49285098819795)(5.185929648241206,39.49239457639077)(5.226130653266332,39.49196717177546)(5.266331658291457,39.49156650293797)(5.306532663316583,39.49119050925271)(5.346733668341709,39.49083731833627)(5.386934673366834,39.49050522622426)(5.42713567839196,39.49019267990547)(5.467336683417085,39.48989826190204)(5.507537688442211,39.48962067662868)(5.547738693467337,39.48935873830316)(5.5879396984924625,39.489111360211574)(5.628140703517588,39.48887754515952)(5.668341708542713,39.48865637696359)(5.708542713567839,39.488447012856994)(5.748743718592965,39.48824867669991)(5.788944723618091,39.488060652899904)(5.8291457286432165,39.487882280959646)(5.869346733668341,39.4877129505797)(5.909547738693467,39.48755209725389)(5.949748743718593,39.4873991983017)(5.989949748743719,39.48725376928951)(6.030150753768845,39.48711536079846)(6.07035175879397,39.48698355550124)(6.110552763819095,39.48685796551506)(6.150753768844221,39.48673823000177)(6.190954773869347,39.4866240129894)(6.231155778894473,39.48651500139224)(6.2713567839195985,39.48641090320941)(6.311557788944723,39.48631144588413)(6.351758793969849,39.48621637480725)(6.391959798994975,39.486125451951594)(6.432160804020101,39.48603845462351)(6.472361809045227,39.485955174321404)(6.5125628140703515,39.48587541569002)(6.552763819095477,39.48579899556243)(6.592964824120603,39.48572574208079)(6.633165829145729,39.48565549388914)(6.673366834170855,39.48558809939144)(6.71356783919598,39.48552341606899)(6.7537688442211055,39.48546130985192)(6.793969849246231,39.485401654540055)(6.834170854271357,39.48534433126873)(6.874371859296483,39.48528922801577)(6.914572864321608,39.48523623914593)(6.954773869346734,39.48518526498972)(6.994974874371859,39.48513621145384)(7.035175879396985,39.48508898966037)(7.075376884422111,39.48504351561237)(7.115577889447236,39.48499970988394)(7.155778894472362,39.48495749733239)(7.1959798994974875,39.48491680683119)(7.236180904522613,39.48487757102168)(7.276381909547739,39.484839726082114)(7.316582914572864,39.48480321151293)(7.35678391959799,39.48476796993672)(7.396984924623116,39.48473394691176)(7.437185929648241,39.48470109075843)(7.477386934673367,39.484669352396956)(7.517587939698492,39.48463868519626)(7.557788944723618,39.48460904483252)(7.597989949748744,39.484580389157266)(7.63819095477387,39.48455267807369)(7.678391959798995,39.48452587342136)(7.71859296482412,39.48449993886785)(7.758793969849246,39.48447483980756)(7.798994974874372,39.484450543266824)(7.839195979899498,39.48442701781472)(7.8793969849246235,39.48440423347981)(7.919597989949749,39.48438216167156)(7.959798994974874,39.484360775106865)(8.0,39.484340047740886)
	};
	\addlegendentry{$L=4$}
	\addplot+[thick, mark=none,smooth]
	coordinates {
		(0.0,0.0)(0.04020100502512563,2.7834886635291607)(0.08040201005025126,7.856431671178168)(0.12060301507537688,11.86037323318587)(0.16080402010050251,14.438087535805488)(0.20100502512562815,16.05704240088757)(0.24120603015075376,17.1037911027357)(0.2814070351758794,17.81008517822467)(0.32160804020100503,18.308400123868378)(0.36180904522613067,18.675867876069)(0.4020100502512563,18.9592185421231)(0.44221105527638194,19.188233180019164)(0.4824120603015075,19.383134587496652)(0.5226130653266332,19.55887354652176)(0.5628140703517588,19.727873230473918)(0.6030150753768845,19.90212135914507)(0.6432160804020101,20.09523618744567)(0.6834170854271356,20.32514515647674)(0.7236180904522613,20.618313348969536)(0.7638190954773869,21.017177076235807)(0.8040201005025126,21.59371212442435)(0.8442211055276382,22.472813463538028)(0.8844221105527639,23.85874402323298)(0.9246231155778895,25.974540876269753)(0.964824120603015,28.51415187864945)(1.0050251256281406,29.605541118913834)(1.0452261306532664,27.946345866976067)(1.085427135678392,25.43767607213756)(1.1256281407035176,23.5460616368688)(1.1658291457286432,22.343893927936932)(1.206030150753769,21.59255687006919)(1.2462311557788945,21.109642244550848)(1.2864321608040201,20.78738971337104)(1.3266331658291457,20.564399388220757)(1.3668341708542713,20.40503586680944)(1.407035175879397,20.287919414945137)(1.4472361809045227,20.199759626761452)(1.4874371859296482,20.132011512857567)(1.5276381909547738,20.07900997202715)(1.5678391959798996,20.0368945900821)(1.6080402010050252,20.00296958664807)(1.6482412060301508,19.975311018855702)(1.6884422110552764,19.95251886615652)(1.728643216080402,19.933556500163622)(1.7688442211055277,19.917644290558997)(1.8090452261306533,19.9041875836755)(1.849246231155779,19.892727002298052)(1.8894472361809045,19.882903542429855)(1.92964824120603,19.87443366606781)(1.9698492462311559,19.867091264724838)(2.0100502512562812,19.860694421369118)(2.050251256281407,19.855095573038504)(2.090452261306533,19.850174116423666)(2.130653266331658,19.845830790578763)(2.170854271356784,19.841983367528837)(2.21105527638191,19.83856331591068)(2.251256281407035,19.835513195854897)(2.291457286432161,19.832784608598896)(2.3316582914572863,19.83033657064985)(2.371859296482412,19.828134215560524)(2.412060301507538,19.82614775048173)(2.4522613065326633,19.82435161229877)(2.492462311557789,19.822723781193588)(2.5326633165829144,19.821245219187464)(2.5728643216080402,19.81989940851578)(2.613065326633166,19.81867197021106)(2.6532663316582914,19.817550347482918)(2.693467336683417,19.81652354171887)(2.7336683417085426,19.815581891430462)(2.7738693467336684,19.814716886413926)(2.814070351758794,19.813921010916445)(2.8542713567839195,19.81318761079656)(2.8944723618090453,19.8125107806144)(2.9346733668341707,19.811885267340845)(2.9748743718592965,19.81130638797676)(3.0150753768844223,19.810769958856675)(3.0552763819095476,19.810272234801296)(3.0954773869346734,19.809809856598854)(3.1356783919597992,19.80937980555241)(3.1758793969849246,19.808979364039743)(3.2160804020100504,19.808606081204676)(3.2562814070351758,19.80825774303999)(3.2964824120603016,19.807932346239145)(3.3366834170854274,19.807628075290573)(3.3768844221105527,19.807343282369047)(3.4170854271356785,19.807076469645523)(3.457286432160804,19.806826273693215)(3.4974874371859297,19.806591451714482)(3.5376884422110555,19.806370869353046)(3.577889447236181,19.806163489889148)(3.6180904522613067,19.805968364643885)(3.658291457286432,19.80578462444262)(3.698492462311558,19.80561147200799)(3.7386934673366836,19.805448175170284)(3.778894472361809,19.805294060797902)(3.819095477386935,19.805148509363114)(3.85929648241206,19.805010950069626)(3.899497487437186,19.804880856477386)(3.9396984924623117,19.80475774256853)(3.979899497487437,19.804641159205058)(4.0201005025125625,19.804530690935106)(4.060301507537688,19.804425953109725)(4.100502512562814,19.804326589276723)(4.14070351758794,19.80423226882211)(4.180904522613066,19.804142684832986)(4.221105527638191,19.804057552158934)(4.261306532663316,19.803976605651357)(4.301507537688442,19.80389959856281)(4.341708542713568,19.80382630109004)(4.381909547738694,19.803756499046614)(4.42211055276382,19.80368999265223)(4.4623115577889445,19.80362659542737)(4.50251256281407,19.803566133183196)(4.542713567839196,19.803508443097513)(4.582914572864322,19.80345337286874)(4.623115577889448,19.803400779940578)(4.663316582914573,19.803350530790723)(4.703517587939698,19.80330250027796)(4.743718592964824,19.803256571042162)(4.78391959798995,19.803212632952512)(4.824120603015076,19.803170582599606)(4.864321608040201,19.80313032282767)(4.9045226130653266,19.803091762303268)(4.944723618090452,19.803054815117374)(4.984924623115578,19.803019400418055)(5.025125628140704,19.802985442070938)(5.065326633165829,19.80295286834538)(5.105527638190955,19.802921611624004)(5.1457286432160805,19.802891608133734)(5.185929648241206,19.802862797696598)(5.226130653266332,19.80283512349864)(5.266331658291457,19.80280853187549)(5.306532663316583,19.80278297211329)(5.346733668341709,19.802758396263755)(5.386934673366834,19.802734758972193)(5.42713567839196,19.802712017317557)(5.467336683417085,19.802690130663596)(5.507537688442211,19.802669060520135)(5.547738693467337,19.80264877041391)(5.5879396984924625,19.802629225768058)(5.628140703517588,19.80261039378969)(5.668341708542713,19.802592243364973)(5.708542713567839,19.802574744961124)(5.748743718592965,19.80255787053481)(5.788944723618091,19.8025415934465)(5.8291457286432165,19.8025258883804)(5.869346733668341,19.802510731269404)(5.909547738693467,19.80249609922492)(5.949748743718593,19.802481970471103)(5.989949748743719,19.80246832428312)(6.030150753768845,19.80245514092938)(6.07035175879397,19.802442401617263)(6.110552763819095,19.802430088442165)(6.150753768844221,19.80241818433966)(6.190954773869347,19.802406673040604)(6.231155778894473,19.802395539028822)(6.2713567839195985,19.802384767501422)(6.311557788944723,19.802374344331422)(6.351758793969849,19.802364256032533)(6.391959798994975,19.802354489726085)(6.432160804020101,19.80234503310977)(6.472361809045227,19.8023358744283)(6.5125628140703515,19.80232700244562)(6.552763819095477,19.8023184064188)(6.592964824120603,19.802310076073365)(6.633165829145729,19.802302001579996)(6.673366834170855,19.802294173532562)(6.71356783919598,19.802286582927337)(6.7537688442211055,19.80227922114338)(6.793969849246231,19.802272079923995)(6.834170854271357,19.802265151359173)(6.874371859296483,19.80225842786902)(6.914572864321608,19.802251902188036)(6.954773869346734,19.802245567350255)(6.994974874371859,19.80223941667516)(7.035175879396985,19.80223344375437)(7.075376884422111,19.802227642439007)(7.115577889447236,19.802222006827684)(7.155778894472362,19.80221653125518)(7.1959798994974875,19.802211210281655)(7.236180904522613,19.80220603868245)(7.276381909547739,19.802201011438342)(7.316582914572864,19.80219612372639)(7.35678391959799,19.802191370911128)(7.396984924623116,19.80218674853631)(7.437185929648241,19.802182252316996)(7.477386934673367,19.80217787813202)(7.517587939698492,19.80217362201689)(7.557788944723618,19.802169480156977)(7.597989949748744,19.802165448881073)(7.63819095477387,19.802161524655222)(7.678391959798995,19.8021577040769)(7.71859296482412,19.802153983869427)(7.758793969849246,19.802150360876638)(7.798994974874372,19.802146832057904)(7.839195979899498,19.802143394483213)(7.8793969849246235,19.80214004532867)(7.919597989949749,19.802136781872054)(7.959798994974874,19.80213360148868)(8.0,19.802130501647415)
	};
	\addlegendentry{$L=2$}
	\end{axis}		
	\end{tikzpicture}
	
	\caption{
		\label{fig:Gamma(t)_AFM_FM}	
		The decoherence function of the off-diagonal matrix element of the density matrix of  $\GHZ$ (top) and $\GHZp$ (bottom) in $d=2$ dimensions, with $\om_c=10/a$ and $T=0$. We observe $L-1$ sharp extrema at $t/a=1,2,\ldots,L-1$. The time interval between these extrema is equal to the time required by the mode to travel a distance $a$. There are no  extrema after the mode has had the time to travel the distance $aL$, which is the total length of the array. For $\GHZp$, the extrema are alternating local maxima and minima. For $\GHZ$, there are only local minima at these points.  After the series of extrema, $\Gamma_\mrm{GHZ}$ and $\Gamma_\mrm{GHZ'}$ reach the same plateaus, the height of which is given by Eq.~(\ref{eq:limit_t_to_infty}). Plots for higher $L$, and odd $L$, show the same behavior.
	}
\end{figure}

Going back to the general case, we have from Eqs.~(\ref{eq:Gamma_explicit}) and (\ref{eq:gamma_computational}), and $f_{\bd 0}=\lVert \bd \rVert^2$, that
\begin{equation}\label{eq:Gamma(t)_pulled_out}
\Gamma^{(vac)}_{\bd}(t)=\alpha_d \lVert \bd \rVert^2 I_0 + \alpha_d \sum_{r=1}^{L-1} f_{\bd r} I_r,
\end{equation} with
\begin{equation}\label{eq:Gamma_as_integral}
I_r(t)=\int_0^\infty \dd\omega\, \omega^{d}\tau(t,\om) \,e^{-\omega/\om_c}\cos(a\om r).
\end{equation}
This integral is solved using standard identities for Gaussian integrals. For $d>0$, $d\neq 1$,
\begin{align}\label{eq:integral_solution_dneq1}
I_{r}(t)=&\frac{a^{1-d}}{4}\tilde\Gamma (d-1) \nn \\
&\times\left[ 2 ( Q_{r0})^{1-d} - ( Q_{r,-1})^{1-d} - (Q_{r1})^{1-d} \right]\nn\\
&+c.c,
\end{align}
Here $c.c.$ stands for the complex conjugate of the preceding term, and 
\begin{equation}
\label{eq:Q_rj} Q_{rj} \defeq i(jt/a-r)+\frac{1}{a\om_c},
\end{equation}
with $i$ the imaginary unit. For $d=1$,
\begin{align}\label{eq:integral_solution_deq1}
I_r(t)&=\frac{1}{4} \left[-2 \log\left(Q_{r0} \right) + \log( Q_{r,-1}) + \log(  Q_{r1})\right]\nn\\
&\phantom{=}+c.c.
\end{align}
We now have $\Gamma_\bd^{(vac)}(t)$ in closed form, except for the sum over a single index in Eq.~(\ref{eq:Gamma(t)_pulled_out}). Using this analytic solution, $\Gamma_\mrm{GHZ}$ and $\Gamma_{\mrm{GHZ}^{'}}$ are plotted in Fig.~\ref{fig:Gamma(t)_AFM_FM}.

\subsection{Infinitesimal time limit}
The leading, second order in time of the integral in Eq.~(\ref{eq:Gamma_as_integral}) equals
\begin{align*}
\frac{1}{2}t^2\int_0^{\infty} \dd \omega\,e^{-\omega/\om_c}\omega^{d}\cos(a\om r).
\end{align*}
Solving this integral, we obtain, for $r=0$,
\begin{equation}\label{eq:I_0_small_times}
I_0(t)=\om_c^{d-1}\left\{ \frac{1}{2}  \tilde\Gamma(1+d) (t\om_c)^2+ O\left[(t\om_c)^4\right]\right\}.
\end{equation}
For $r>0$, we find
\begin{align*}
I_{r>0}(t)&=a^{1-d}\frac{1}{4}\tilde \Gamma(1+d)(t/a)^2[(Q_{r0})^{-(d+1)}+c.c.]\\ 
&\phantom{=}+a^{1-d}O\left[(t/a)^4\right].
\end{align*}
These two solutions hold for all $d>0$. Up to a factor $\alpha_d \lVert \bd \rVert^2$, the first term of the decoherence function in Eq.~(\ref{eq:Gamma(t)_pulled_out}) is given by Eq.~(\ref{eq:I_0_small_times}). For the remaining terms, with $r>0$, note that $|f_{\bd r}|\leq2 f_{\bd 0}=2\lVert\bd\rVert^2$. Thus, 
\begin{equation}\label{eq:small_times_remaining_terms}
\alpha_d \left|\sum_{r=1}^{L-1} f_{\bd r} I_r(t)\right| \leq 2 \alpha_d \lVert \bd \rVert^2 \sum_{r=1}^{L-1} | I_r(t)|.
\end{equation}
Note $| I_r(t)|$ is proportional to  
\begin{align*}
\left|(Q_{r0})^{-(d+1)}+c.c.\right|&< 2 |Q_{r0}|^{-(d+1)} \\
&=2\left(r^2+\frac{1}{(a\om_c)^2}\right)^{-(d+1)}\\
&<2\,r^{-2(d+1)}    .
\end{align*}
Therefore,
\begin{align*}
\sum_{r=1}^{L-1}\left|(Q_{r0})^{-(d+1)}+c.c.\right|&<2 \sum_{r=1}^{L-1} r^{-2(d+1)} \\
&< 4.
\end{align*}
Thus, we obtain
\begin{align}
\alpha_d\left|\sum_{r=1}^{L-1} f_{\bd r} I_r\right| &< 2 \alpha_d \lVert \bd \rVert^2 a^{1-d}  \tilde \Gamma(1+d) (t/a)^2 \nn\\
&\phantom{<}+ \alpha_d \lVert \bd \rVert^2 a^{1-d} O(t/a)^4. \label{eq:short_times_remaining_terms_2}
\end{align}
There are extra factors of $L$ hiding in the $O(t/a)^4$ term. We can disregard this $L$ dependence because, in this subsection, we are interested in the limit of infinitesimal time. Then for any $L$ there is a $t/a\ll1$ such that the second term in Eq.~(\ref{eq:short_times_remaining_terms_2}) is negligible.

Thus, for small times and $d>0$, the final result is
\begin{align}
\Gamma^{(vac)}_\bd(t)=\half \alpha_d \lVert \bd \rVert^2 \tilde \Gamma(1+d) \om_c^{d-1}(t\om_c)^2 + \mc E,
\end{align}
where $\mc E$ contains both the error from the $\alpha \lVert \bd \rVert^2 I_0$ term, and all of the remaining terms in Eq.  (\ref{eq:Gamma(t)_pulled_out}),
\begin{align*}
\mc E & =  \alpha_d \lVert \bd \rVert^2  \om_c^{d-1} O(t\om_c)^4 \\
& \phantom{=}+2 \alpha_d \lVert \bd \rVert^2 a^{1-d}  \tilde \Gamma(1+d) (t/a)^2 \\
& \phantom{=} + \alpha_d \lVert \bd \rVert^2 a^{1-d} O(t/a)^4.
\end{align*}
Given an $L$ and $\om_c a\gg 1$, the relative error
\begin{align*}
\tilde{\mc E}&\defeq \frac{\mc E}{\half \alpha_d \lVert \bd \rVert^2 \tilde \Gamma(1+d) \om_c^{d-1}(t\om_c)^2}\\
&<O(t\om_c)^2 + \frac{4}{(a\om_c)^{d+1}}+ \frac{1}{a\om_c (t\om_c)^2}O(t/a)^4,
\end{align*}
is negligible for $t$ small compared to $1/\om_c$ and $a$.

\subsection{Infinite time limit}\label{sec:infinite_time_limit_appendix}
If $d>1$ and $j\neq 0$,  the function  $(Q_{rj})^{1-d}$ vanishes in the limit that $t$ goes to infinity. For $j=0$, on the other hand, $(Q_{rj})^{1-d}$ is time-independent and nonzero. Thus, from Eq.~(\ref{eq:Gamma_explicit}),
\begin{align}\label{eq:limit_of_I_r}
\lim_{t\to \infty} I_r(t) &=\half a^{1-d}  \tilde \Gamma(d-1) (Q_{r0})^{1-d}+c.c,
\end{align}
for $d>1$. Therefore, $\lim_{t\to\infty}\Gamma^{(vac)}_\bd(t)$ exists for $d>1$, and its value can be found by substituting Eq.~(\ref{eq:limit_of_I_r}) into Eq.~(\ref{eq:Gamma(t)_pulled_out}). The existence of this limit means the vacuum decoherence function always reaches a proper plateau for $d>1$ (cf. Sec. \ref{sec:infinite_time_limit}).

We now show the height of this plateau scales linearly with $L$ for $d\geq 2$, and, for these $d$, simplify the exact expression for the height of the plateau. (This result need not imply superlinear scaling of the height of the plateau for $d<2$.) Firstly,
\begin{equation}\label{eq:limit_of_Gamma}
\lim_{t\to\infty} \Gamma_{\bd}(t) = \alpha_d \sum_{r=0}^{L-1} f_{\bd r}  \lim_{t\to \infty} I_r.
\end{equation}
With $f_{\bd 0}{=}  \lVert\bd\rVert^2$, Eq.~(\ref{eq:limit_of_I_r}), and $Q_{00}=1/(a\om_c)$, the first term ($r=0$) equals
\begin{equation*}
\alpha_d \lVert\bd\rVert^2 \lim_{t\to\infty} I_0 {=}\alpha_d \lVert\bd\rVert^2 \tilde \Gamma(d-1) \om_c^{d-1}.
\end{equation*}
The remaining terms in Eq.~(\ref{eq:limit_of_Gamma}) can be neglected. This is because they are upper bounded by
\begin{align}
\mc E \defeq \alpha_d\left| \sum_{r=1}^{L-1} f_{\bd r}\lim_{t\to\infty} I_r \right| &< 2\alpha_d\lVert\bd\rVert^2  \sum_{r=1}^{L-1} \left| \lim_{t\to\infty} I_r \right|&\nn \\
& \leq 2\alpha_d\lVert\bd\rVert^2 \half a^{1-d} \tilde \Gamma(d-1) &\nn  \\
&\phantom{=} \times \sum_{r=1}^{L-1}\left| (Q_{r0})^{1-d} + c.c.\label{eq:d>2remaining} \right|.&\end{align}
For $r\geq 1$, $|Q_{r0}|{>}1$, and $d\geq 2$, we have 
\begin{align}
\left|(Q_{r0})^{1-d}+c.c.\right|&\leq\left|(Q_{r0})^{-1}+c.c.\right|\label{eq:d=2bound}\\
&=\frac{1}{a\om_c}\frac{1}{r^2+\frac{1}{(a\om_c)^2}}\nn\\
&< \frac{1}{a\om_c}\frac{1}{r^2}.\nn
\end{align}
Thus, with $\sum_{r=1}^{L-1} 1/r^2 <2$, we have for the sum in Eq.~(\ref{eq:d>2remaining}) that
\begin{align*}
\sum_{r=1}^{L-1}\left| (Q_{r0})^{1-d} + c.c. \right|< \frac{2}{a\om_c}.
\end{align*}
Therefore 
\begin{equation*}
\mc E <2\alpha_d \lVert\bd\rVert^2 a^{1-d}\tilde \Gamma(d-1)\frac{1}{a\om_c}.
\end{equation*}
In conclusion, we have for $d\geq 2$,
\begin{equation}\label{eq:limit_t_to_infty}
\lim_{t\to\infty} \Gamma_{\bd}^{(vac)}(t)=\alpha_d \lVert\bd\rVert^2 \tilde\Gamma{(d-1)}\om_c^{d-1}+\mc E,
\end{equation}
with relative error
\begin{align*}
\tilde{\mathcal{E}} &\defeq \frac{ \mc{E}}  {\alpha_d\lVert\bd\rVert^2\tilde\Gamma{(d-1)}\om_c^{d-1}}\\
&<\frac{2}{(a\om_c)^d}\\
&<\frac{2}{a\om_c}.
\end{align*}
The latter is negligible for $a\om_c \gg 1$. Note that $\Gamma_{\bd}^{(vac)}(t) = O(L)$ even if this condition does not hold. 

\section{Dynamical fidelity susceptibility of dephasing}\label{sec:dynamical_fidelity_susceptibility_of_pure_dephasing}
In Ref. \cite{kattemolle2019dynamical}, we studied the leading order effect of an perturbation to the system-reservoir coupling on states in a decoherence-free subspace. We defined this leading order as the dynamical fidelity susceptibility of decoherence-free subspaces. In this previous work we did not assume a particular Hamiltonian, in contrast to the current work, where we focus on the single-reservoir dephasing Hamiltonian. So, on the one hand, the previous work applies more generally. On the other hand, in the previous work we assumed a pure initial reservoir state, whereas in the current work, we assume a product of displaced thermal states. These two assumptions on the initial reservoir state describe different situations. It is only for the vacuum state that both assumptions are simultaneously satisfied. 

In this appendix, we compute the dynamical fidelity susceptibility of decoherence-free subspaces in the specific case that the Hamiltonian is given by the pure single-reservoir dephasing Hamiltonian. This is done in two ways: first by using the general result from our previous work, and then by a more direct computation that does not require the results of the previous work. The expressions we find are identical. However, there is one subtle difference: in the former method, the expression contains the expectation value of the number operator, $\langle N_\bk \rangle_\varphi$, with respect to the pure initial reservoir state $\ket{\varphi}$ (in agreement with the assumptions in the previous work). In the latter method, the same expression contains instead the expectation value of the number operator with respect to the thermal part of a product of displaced thermal states $\bar N_\bk$ (in agreement with the assumptions in the current work). Naturally, the two expressions agree on the subset of initial reservoir states that satisfy both the assumptions on the reservoir states in the current and the previous work. 

We now briefly introduce the result of our previous work, and sequentially use this result to derive the dynamical fidelity susceptibility of the single-reservoir spin-boson dephasing Hamiltonian. Consider a quantum register that is coupled to a reservoir. In general, the overall Hamiltonian is  of the form $H_0=H_S\otimes \id +\id \otimes H_B + H_{SB}$. Here $H_S$ is the system Hamiltonian, $H_B$ the reservoir Hamiltonian, and $H_{SB}$ the interaction term.  Assume that the initial overall state is a product state between the system and the reservoir, $\ket\Psi=\ket\psi\otimes\ket\varphi$, with $\ket\psi$ the initial system state and $\ket \varphi$ the initial reservoir state. 
A decoherence-free subspace (DFS) is a subspace of the system's Hilbert space that does not entangle with the reservoir as $\ket \Psi$ is evolved under the Hamiltonian $H_0$, despite the coupling $H_{SB}$. States in a DFS only experience the unitary evolution due to $H_S$. That is, by definition of a DFS, we may write the system state, in the Schr\"odinger picture, as 
\begin{align}
\rho_0^{Sch}(t)&\defeq \tr_B\left(e^{-itH_0}\ket\psi\ket{\varphi}\!\bra{\psi}\bra{\varphi}e^{itH_0}\right)\nn\\
&=e^{-itH_S}\ket\psi\bra{\psi}e^{itH_S}\nn\\
&=:\ket{\psi(t)}^{Sch}\bra{\psi(t)}^{Sch}.\label{eq:rho_sch}
\end{align}
We write a superscript `$Sch$' when we are specifically referring to Schr\"odinger picture states. 

In Ref. \cite{kattemolle2019dynamical} we compared the system state $\rho^{Sch}_0(t)$ to the state 
\begin{align}
\rho^{Sch}_\ep(t)&\defeq \tr_B\left(e^{-itH_\ep}\ket\psi\ket{\varphi}\!\bra{\psi}\bra{\varphi}e^{itH_\ep}\right), \label{eq:rho_perturbed}
\end{align}
where $H_\ep$ contains an extra interaction term $\ep V$;
\begin{equation}\label{eq:H_perturbed}
H_\ep=H_0+\ep V,
\end{equation}
with $\ep\ll 1$. In general $V$ may be written as  
\begin{equation*}
V=\sum_\alpha S_\alpha\otimes B_\alpha. 
\end{equation*}

To compare the state (\ref{eq:rho_sch}) to (\ref{eq:rho_perturbed}), we computed the dynamical fidelity, which is defined as the fidelity between $\rho^{Sch}_0(t)$ and $\rho^{Sch}_\ep(t)$. Because the former state remains pure, this fidelity has the simple form
\begin{equation}\label{eq:fidelity}
F\left[\rho^{Sch}_0(t),\rho^{Sch}_\ep(t)\right]=\bra{\psi(t)}^{Sch} \rho^{Sch}_\ep(t) \ket{\psi(t)}^{Sch}.
\end{equation}
In the interaction picture, where $H_0$ is the bare Hamiltonian, and $\ep V$ the interaction term, it is straightforward to show that
\begin{align}
F\left[\rho^{Sch}_0(t),\rho^{Sch}_\ep(t)\right]&=F\left[\rho_0^{Int}(t),\rho^{Int}_\ep(t)\right]\nn\\
&=\bra{\psi} \rho_\ep^{Int}(t) \ket{\psi}.\label{eq:fidelity_interaction}
\end{align}
Here $\ket\psi$ is the initial system state, and $\rho_\ep^{Int}(t)$ the reduced system state in the interaction picture,  evolved in time through the interaction-picture time-evolution operator, with $\ep\neq 0$.

The leading order of $F$ in both $t$ and $\ep$ is the second. We defined this leading order as the dynamical fidelity susceptibility of decoherence-free subspaces,
\begin{equation}\label{eq:def_chi}
\chi \defeq  \left. -\frac{1}{4} \frac{\del^2 }{\del \ep^2}\frac{\del^2 F}{\del t^2}\right|_{\ep=t=0}.
\end{equation}
This quantifies the leading order effect of the added system-reservoir coupling on states in a DFS. We have shown that
\begin{equation}\label{eq:previous_work}
\chi=\sum_{\alpha\beta}\langle B^\dagger_\al B^\nodagger_\be\rangle_\varphi^\nodagger[\langle S^{\dagger}_\al S_\be^{\nodagger}\rangle_\psi^\nodagger-\langle S^\dagger_\alpha \rangle_\psi^\nodagger \langle S_\be^{\nodagger}\rangle_\psi^\nodagger],
\end{equation}
where the expectation values are with respect to the initial system and initial reservoir state. This equation is not specific to the single-reservoir dephasing model, but holds in general. 

Using this general result, we can compute the dynamical fidelity susceptibility of the spin-boson single-reservoir dephasing model. After making the substitution {${g_{\bk}\to\ep g_\bk}$}, the Hamiltonian of the single-reservoir dephasing model [Eq.~(\ref{eq:collective_pure_dephasing})] is of the form of Eq.~(\ref{eq:H_perturbed}), with $H_{SB}=0$ and 
\begin{align*}
S_\ell=J^z_\ell, &&B_\ell=\sum_{\bk} (g_{\bk \ell}^* a_\bk^{\nodagger}+g_{\bk\ell}^\nodagger a_\bk^\dagger).
\end{align*}
Note that, because $H_{SB}=0$, actually the entire Hilbert space of the system is a DFS. For system states of the form $\ket\psi=(\ket\bi+\ket\bj)/\sqrt{2}$, we find
\begin{align*}
\chi=&\,\frac{1}{4}\sum_{\ell m \bk \bk'} \bd_\ell^\nodagger \bd_m^\nodagger \left[g_{\bk\ell}^{\phantom{*}}g_{\bk' m}^*(\delta_{\bk\bk'}^\nodagger+2\langle a_\bk^\dagger a_{\bk'}^\nodagger \rangle_\varphi) \right. \\ 
& \left.+\,g_{\bk \ell}^\nodagger g_{\bk'm}^\nodagger \langle a_\bk^\dagger a_{\bk'}^{\dagger}\rangle_\varphi + c.c. \right],
\end{align*}
with $\bd=\bi-\bj$, and where $c.c.$ stands for the complex conjugate of the preceding term only. If the initial reservoir state $\ket\varphi$ is a product of number states, then  
\begin{equation*}
\chi=\frac{1}{4}\sum_{\ell m \bk} \bd_\ell^\nodagger \bd_m^\nodagger g_{\bk\ell}^\nodagger g_{\bk m}^*(1+2\langle N_{\bk}^\nodagger \rangle_\varphi).
\end{equation*}
Using $g_{\bk\ell}=g_\bk e^{i\bk\cdot\br_\ell}$ and writing $\gamma_\bd(\bk)$ as the spectral density of $\bd$ (see Sec. \ref{sec:dephasing_susceptibility}), we obtain the result
\begin{equation}\label{eq:chi_dynamical_fidelity}
\chi=\frac{1}{4}\sum_{\bk}\lvert g_\bk \rvert^2\gamma_\bd(\bk)(1+2\langle N_{\bk}\rangle_\varphi).
\end{equation}

Note that we have now obtained $\chi$ without ever solving for the reduced time evolution of the system state. This illustrates that Eq.~(\ref{eq:previous_work}) can be used to study superdecoherence in models that have not been solved. The condition for superdecoherence would read $\chi\propto L^2$ instead of $\Gamma_L\propto L^2$. One caveat is that $\chi$ is a leading order in time. With Eq.~(\ref{eq:chi_dynamical_fidelity}) we have already obtained a result that could not be obtained by using the full time evolution of the spin-boson single-reservoir dephasing model, because the latter relies on the assumption that the initial reservoir state is a displaced thermal state. Equation~(\ref{eq:chi_dynamical_fidelity}) holds for general pure initial reservoirs states, some of which cannot be described as a displaced thermal state (with $T=0$). Furthermore, we can take the continuum limit of Eq.~(\ref{eq:chi_dynamical_fidelity}), just as we have done in the main text for $\Gamma_L$, and show, in exactly the same way, that $\chi=O(L)$ in continuous reservoirs. This extends the results in the main text to include arbitrary pure reservoir states, albeit only for the leading order in time.

For the single-reservoir dephasing model, we actually have the full solution of the reduced system density matrix at hand, which means the dynamical fidelity susceptibility may be obtained by other means. We may solve for the interaction-picture states $\rho^{Int}_0(t)$ and $\rho^{Int}_\ep(t)$, and compute derivatives of Eq.~(\ref{eq:fidelity}). In the main text, are interested in the absolute value of the system density matrix only. Here, however, we need the full solution, which, in the interaction-picture, reads \cite{reina2002decoherence}
\begin{equation*}
\rho^{Int}_{\bi\bj}(t)=e^{i[\Theta_{\bi\bj}(t)-\Lambda_{\bi\bj}(t)]}e^{-\Gamma_{\bi-\bj}(t)}\,\rho^{Int}_{\bi\bj}(0),
\end{equation*}
where 
\begin{align*}
\Theta_{\bi\bj}(t)=&\sum_{\bk}\lvert g_\bk \rvert^2  \frac{\om_\bk t - \sin(\om_\bk t)}{\om_\bk^2} \\
&\times  \sum_{\ell m}(\bi_\ell \bi_m - \bj_\ell \bj_m)\cos(\bk\cdot \br_{\ell m}),
\end{align*}
and
\begin{equation*}
\La_{\bi\bj}(t)= 2 \sum_\bk \lvert g_\bk \rvert^2 \tau(\om_\bk,t) \sum_{\ell m}\bi_\ell \bj_m \sin (\bk\cdot\br_{\ell m}).
\end{equation*}
This is not in the continuum limit.

With $\ket{\psi}=\sum_\bi \psi_\bi \ket \bi$, and using the solution for $\rho^{Int}_{\bi\bj}(t)$, the dynamical fidelity [Eq.~(\ref{eq:fidelity_interaction})] reads
\begin{align}
F\left[\rho^{Int}_0(t),\rho^{Int}_\ep(t)\right]=&\sum_{\bi\bj}\psi_{\bi}^*\rho_{\bi\bj}(t)\psi_{\bj}\nn\\
=&
\sum_{\bi\bj} |\psi_{\bi}|^2
|\psi_{\bj}|^2 \nn\\
&\times e^{i\left[\Theta_{\bi\bj}(t)-\Lambda_{\bi\bj}(t)\right]}e^{-\Gamma_{\bi-\bj}(t)}. \label{eq:pure_dephasing_fidelity}
\end{align}
This expression is real because of the antisymmetry of $\Theta_{\bi\bj}(t)-\Lambda_{\bi\bj}(t)$ in $\bi$ and $\bj$, and because $\Theta_{\bi\bi}(t)=\La_{\bi\bi}(t)= 0$. For states of the form $\ket \psi = (\ket{\bi}+\ket{\bj})/\sqrt{2}$, Eq.~(\ref{eq:pure_dephasing_fidelity}) simplifies to
\begin{align}
F&=\half + \half \cos\left[\Theta_{\bi\bj}(t)-\Lambda_{\bi\bj}(t)\right]\, e^{-\Gamma_{\bi_0-\bj_0}(t)}.
\end{align}
In special cases, the cosine equals unity for all $t$. Examples  include $(\bi,\bj)=(\bi_\mrm{GHZ},\bj_\mrm{GHZ})$ and $(\bi,\bj)=(\bi_{\mrm{GHZ}^{'}},\bj_{\mrm{GHZ}^{'}})$. Computing the derivatives of $F$ [see Eq. (\ref{eq:def_chi})], we obtain
\begin{align}
\chi&=\frac{1}{4}\Gamma^{''}_\bd (0)\nn\\
&=\frac{1}{4}\sum_\bk\lvert g_\bk \rvert^2 \gamma_{\bd}(\bk) (1+2\bar N_\bk).\label{eq:chi_pure_dephasing}
\end{align}

Note the similarity between Eqs.~(\ref{eq:chi_dynamical_fidelity}) and (\ref{eq:chi_pure_dephasing}). The assumptions involved in deriving these two equations, however, are differexnt. In the former, the assumption is that the initial reservoir state is pure. In the latter, it is assumed that the reservoir state is a product of unentangled displaced thermal states (see Sec.~\ref{sec:pure_dephasing}). The only reservoir state for which these two assumptions coincide is the vacuum state. In this case, both expressions agree, and read
\begin{equation*}
\chi=\frac{1}{4}\sum_\bk\lvert g_\bk \rvert^2 \gamma_{\bd}(\bk).
\end{equation*}
The similarity of Eqs.~(\ref{eq:chi_dynamical_fidelity}) and (\ref{eq:chi_pure_dephasing}) indicates that our assumptions on the initial reservoir state may be relaxed in both situations.

\end{document}